\documentclass[11pt]{amsart}
\usepackage{geometry}                % See geometry.pdf to learn the layout options. There are lots.
\usepackage[parfill]{parskip}    % Activate to begin paragraphs with an empty line rather than an indent
\usepackage{graphicx}
\usepackage{amssymb}
\usepackage{mathtools}
\usepackage{epstopdf}
\usepackage{enumitem}
\usepackage{xcolor}
\usepackage[utf8]{inputenc}
\usepackage[T1]{fontenc}
\usepackage{url}   %% used in references
\usepackage{ulem}

\usepackage{hyperref}

\usepackage[utf8]{inputenc}
\usepackage[english]{babel}
 
\newtheorem{theorem}{Theorem}[section]
\newtheorem{corollary}{Corollary}[section]
\newtheorem{claim}{Claim}[section]
\newtheorem{lemma}[theorem]{Lemma}
\newtheorem{definition}{definition}[section]
\newtheorem{remark}{remark}[section]

\newcommand{\bibref}[1]{Ref.~\cite{#1}}
\renewcommand{\eqref}[1]{Eq.~(\ref{#1})}

\renewcommand{\d}{{\rm d}}

\title{\hspace{24pt} Factorization of Linear Quantum Systems \newline with Delayed Feedback}
\author{
  Gil Tabak
  \and
  Ryan Hamerly
  \and
  Hideo Mabuchi
}
\begin{document}
\maketitle

%TODO: Re-write intro and add abstract with good physical interpretation.

% \red{TODO: add all references what else to cite: Hendra Nurdin's work, several books about J-unitary stuff,}

%TODO: Insert proof for Ryan's theorems in Section~\ref{sec:special_cases}.

%TODO: Discuss static squeezing

%Question (interesting, but not important for this paper): Can we do the same in Section~\ref{sec:another_form} for real roots?

%Misc: which notation is best for $J$-unitary and doubled-up matrices in this paper: alternating port types (creation and annihilation), or first all annihilation and then all creation?

\begin{abstract}
We consider the transfer functions describing the input-output relation for a class of linear open quantum systems involving feedback with nonzero time delays.
We show how such transfer functions can be factorized into a product of terms which are transfer functions of canonical physically realizable components.
We prove under certain conditions that this product converges, and can be approximated on compact sets.
Thus our factorization can be interpreted as a (possibly infinite) cascade.
Our result extends past work where linear open quantum systems with a state-space realization have been shown to have a pure cascade realization [Nurdin, H. I., Grivopoulos, S., \& Petersen, I. R. (2016). The transfer function of generic linear quantum stochastic systems has a pure cascade realization. Automatica, 69, 324-333.].
The  functions we consider are inherently non-Markovian, which is why in our case the resulting product may have infinitely many terms. 
\end{abstract}

\section{Overview}

\label{sec:overview}

%\red{Moved to the beginning (this is your introductory section?).  TODO: motivate the result.  Reference other factorization, delay papers, including your own, Nurdin's cascade synthesis papers, probably also Grimsmo et al. PRL 115, 060402.  Hideo might be able to help.}

We will be concerned here with a class of non-Markovian linear open quantum systems. 
Specifically, we will construct this class of systems by adding delayed feedback 
to a (Markovian) finite-dimensional open quantum system.
After this procedure, the transfer function may not correspond to a rational function.
We will show that under certain conditions the input-output relationship of such systems can be realized as a (possibly infinite) cascade of canonical components. 
To do this, we will factorize the transfer function into an infinite product of canonical terms, where each term corresponds to a physically realizable component. Each such term will be identified using a pair of poles and zeros.
The delayed-feedback system can then be approximated using a finite truncation of the infinite product. One application of our result is providing experimentalists with a method to generate a cascade network of canonical terms replicating the behavior of a system with delayed feedback.

Our work provides an extension of some prior papers, and is closely related to others. In \cite{grivopoulos2016transfer} and \cite{nurdin2016transfer}, the authors obtain a cascade realization from a given linear quantum stochastic system. Our work here can be considered a natural extension in the case of delayed feedback.  Isolated loops with time delays are also discussed in the context of linear quantum systems in \cite{gough2017isolated}. In \cite{tabak2016trapped}, a cascade realization is given for passive linear systems with delayed feedback. In that case, it suffices to consider only the input-output relationship for the annihilation operators. 
In the current paper we consider more generally quantum linear systems which may not be passive.  In order to characterize the transfer function for such a system, we must consider the input-output relationship between both the creation and annihilation operators, since the system may have squeezing effects. 

There are also approaches that have been developed for simulating coherent feedback systems with time delays for the case when the system may be in general nonlinear.
In \cite{grimsmo2015time}, a series of cascaded systems is introduced, such that the system is driven by past versions of itself. Later in \cite{whalen2017open}, this work was extended to include the case of multiple delays of different durations. In \cite{pichler2016photonic}, matrix product states were used study the dynamics of coherent feedback networks with significant delays.   These approaches  differ from our approach where a cascade of canonical terms is constructed in the frequency domain (which is well-defined for linear systems) by considering the shape of the transfer function near the zeros and poles. 

In the case of rational transfer functions, properties of the the state-space representation matrices can be used to determine if a system is physically realizable. It has been shown in ~\cite{shaiju2012frequency} that under some non-degeneracy assumptions these conditions are equivalent to certain properties of the transfer function in the frequency domain. Essentially, as long as the state-space realization matrices are in `doubled-up' form and certain non-degeneracy conditions hold, the physical realizability of the system is equivalent to the transfer function having the $J$-unitary property (and a property preventing static squeezing).
Once delayed feedback is incorporated, although the system remains linear, it loses its Markovian property and no longer has a corresponding state-space representation. 
We introduce an extension of the frequency domain constraints for physical realizability and utilize it to obtain our factorization under certain assumptions, stated in Section~\ref{sec:assumptions} and in the non-degeneracy conditions in the hypothesis of Theorems~\ref{theorem:complex_case} and~\ref{theorem:real_case}. Some of the assumptions follow from our construction starting with a finite-dimensional physically realizable system (Assumptions~\ref{assumption:finite_dim} and~\ref{assumption:physically_realizable}), and others are introduced for simplicity (Assumptions~\ref{assumption:commensurate},~\ref{assumption:zeros}, and~\ref{assumption:simple}). Assumption~\ref{assumption:proper} ensures the network does not have effectively time-delayed feed-forward terms. An analysis similar to the one in~\cite{tabak2016trapped} can be used to isolate such terms, but this is outside the scope of our paper. 

%The goal will ultimately be to describe how this can be done for (1) general linear systems, (2) passive (energy conserving) systems, (3) active linear quantum systems (ALQ) with or without static squeezing. The condition for (2) is that the transfer function is unitary along the imaginary axis. For (3), the transfer function  considered will be in the `doubled-up' notation. The condition for (3) is that the transfer function is $J$-unitary. No static squeezing for a rational matrix valued function represented by a doubled-up transfer function $T(z) = D + C(zI - A)^{-1}B$ implies that the $D$ matrix of the doubled-up transfer function has form $D = \Delta(S,0)$. If the transfer function is J-unitary along the imaginary axis and has no static squeezing, I will call it physically realizable (see~\cite{shaiju2012frequency}).

%\red{This is a long paper.  The last paragraph of the introduction should map out its structure, e.g. ``In Sec.~2 we do \ldots, then in Sec.~3 we do \ldots, blah blah\ldots''.  Table of contents wouldn't hurt either, if it's allowed by the journal.}

In Section~\ref{sec:def}, we introduce the various definitions and notation we use throughout.
In Section~\ref{sec:LQSS} we introduce (finite-dimensional) linear quantum stochastic systems and discuss the physical realizability conditions of transfer functions. In Section~\ref{sec:delayed_feedback} we construct a delayed-feedback network and state assumptions we make throughout. In Section~\ref{sec:TF_claims} we provide some properties of our delayed-feedback network. In Section~\ref{section:static} we provide theorems that relate the transfer function of the delayed-feedback network with a simpler transfer function resulting from replacing the internal components of the network with static components. These theorems will be useful in providing convergence results. In Section~\ref{sec:fundamental_factors} we introduce physically realizable canonical transfer functions having only two poles and two zeros.  Finally, in Section~\ref{sec:fac_QLS}, we provide our factorization result.

\section{Definitions}
\label{sec:def}

%TODO: Organize definitions and overview

%\section{\blue{\sout{Matrices and matrix-valued functions}}}

%\red{These are from a James / Nurdin / Petersen paper?  Cite so readers are familiar.}
{The definitions below are frequently used in the literature (e.g.~\cite{gough2010squeezing},\cite{gough2009enhancement},\cite{shaiju2012frequency}).}

\begin{definition}
The signature matrix $J$ of dimension $2n$ is a diagonal matrix with alternating $(+1,-1)$ on its diagonal.
\end{definition}

\begin{definition}
\label{def:flat}
For a matrix $M$ we use the notation $M^\flat = J M^\# J$.
\end{definition}

\begin{definition}
\label{def:Delta}
{A doubled-up matrix has the form
\begin{align}
\breve{M} = \Delta (M_-, M_+)
 \equiv  \begin{pmatrix}
 M_- && M_+ \\
 M_+^\# && M_- ^\#
 \end{pmatrix}. \label{eq:dblmat}
\end{align}
A doubled-up column vector has the form
\begin{align}
\breve{v} = \begin{pmatrix}
 v \\
 v ^\#
 \end{pmatrix}. \label{eq:dblvec}
\end{align}
Here $M^\#, v^\#$ denote the elementwise conjugates of $M, v$}.
\end{definition}

\begin{definition}
A matrix $M$ is $J$-unitary iff
\begin{align}
MJM^\dag = M^\dag J M  = J.
\end{align}
\end{definition}

\begin{definition}
A matrix-valued function $M(z)$ is $J$-unitary iff
\begin{align}
M(z) J M(z)^\dag = M(z)^\dag J M(z) = J,
\end{align}
for z on the imaginary axis.
\end{definition}

It a transfer function is $J$-unitary and meromorphic on the plane, then satisfies 
$M(z) J  M(- \overline z)^\dag =J$ on its domain. 

It will be convenient to permute the annihilation and creation operators using the symmetric permutation matrix $P$, exchanging the indices the first $N$ indices, $1,2,...,N$ with the odd indices up to $2N$, and exchanging  the last $N$ indices $N+1,...,2N$ with the even indices up to $2N$. When this is the case (Section~\ref{sec:delayed_feedback} and beyond) it will be understood that doubled-up matrices will have the form $\tilde M = P M P^T$ instead. % \red{A bit confusing, though I know what you mean.  Maybe give an example with a simple matrix?}
For example, the matrix
\begin{align}M  = \Delta(M_-, M_+) = 
	\begin{pmatrix}
		M_{-,11}  & M_{-,12}  & M_{+,11}  & M_{+,12} \\
		M_{-,21}  & M_{-,22} & M_{+,21}  & M_{+,22} \\
		M_{+,11} ^\# & M_{+,12}^\# & M_{-,11}^\#  & M_{-,12} ^\# \\
		M_{+,21}^\#  & M_{+,22} ^\# &M_{-,21}^\#  & M_{-,22} ^\#
	\end{pmatrix}
\end{align}
would be re-arranged as
\begin{align}P M P^T = 
	\begin{pmatrix}
		M_{-,11}  & M_{+,11} &  M_{-,12}  & M_{+,12} \\
		M_{+,11} ^\# & M_{-,11} ^\# & M_{+,12} ^\# & M_{-,12} ^\# \\
		M_{-,21}  & M_{+,21} &  M_{-,22}  & M_{+,22} \\
		M_{+,21} ^\# & M_{-,21} ^\# & M_{+,22} ^\# & M_{-,22} ^\# \\
	\end{pmatrix}.
\end{align}
\begin{definition}
We will say a rational transfer function is in `doubled-up' form when it has a realization given by matrices $A,B,C,D$, (i.e. $T(z) = D + C(zI - A)^{-1}B$), where each is in doubled-up form.
\end{definition}

\begin{definition}
Given a dimension $b$, let
 \begin{align}
\Sigma_b = \begin{pmatrix}
0 && I_b \\ I_b && 0
\end{pmatrix} 
\end{align}
The subscript $b$ will be dropped when the dimension is clear from context.
\end{definition}
It will be understood that when the annihilation and creation ports are permuted as discussed above,
we will use $\tilde \Sigma = P \Sigma P^T$ instead.

\begin{remark}
A matrix $M$ is doubled-up if and only if it satisfies
\begin{align}
M \Sigma = \Sigma M^\#,
\end{align}
\end{remark}

\subsection{Zeros and poles}

In the literature, there are different definitions for zeros and poles of a matrix-valued function. We will describe two kinds of definitions, which we will denote respectively by (I) and (II):

\begin{definition}
\label{def:zeros_I}
A complex number z is called a pole(I) of $A(z)$ if it is a pole of one of the entries of $A(z)$, and z is called a zero(I) of $A(z)$ if it is a pole(I) of $A(z)^{-1}$. This is the definition used in  \bibref{prather1987factorization}.
\end{definition}

\begin{definition}
\label{def:zeros_II}
A complex number z is called a pole(II) of $A(z)$ if it is a pole of $\det A(z)$, and z is called a zero(II) of $A(z)$ if it is a zero of $\det A(z)$.  This is the definition used in \bibref{tabak2016trapped}.  Note that definition \ref{def:zeros_II} implies \ref{def:zeros_I}, but not the converse.
\end{definition}

\begin{remark}
The definitions (I) could have a pole and a zero in the same location. Instead, for definitions (II) this cannot occur. If zeros (I) and poles (I) do not occur in the same location, then the definitions (I) and (II) coincide. For simplicity, we will assume that is the case (Assumption~\ref{assumption:zeros}).
\end{remark}

\begin{definition} 
\label{def:eig_vectors}
(eigenvectors at zeros and poles, adapted from~\bibref{prather1987factorization})
Suppose $z_0$ is a zero(I) of matrix valued function $A(z)$. A nonzero column vector $x_1$ is called an eigenvector of $A(z)$ at $z_0$ if there exist column vectors $(x_2,x_3,\ldots)$ such that $f(z) = A(z) \sum_{j=0}^\infty x_{j+1} (z - z_0)^j$ is analytic at $z_0$ and has a zero at $z_0$. We will say the order of the zero of $A(z)$ with eigenvector $x_1$ at $z_0$ is the order of $f(z)$ at $z_0$.

Similarly, if $z_0$ is a pole of $A(z)$, then $y_1$ is a pole vector of $A(z)$ at $z_0$ (we will sometimes say an eigenvector at the pole)  if there exist vectors $(y_2,y_3,\ldots)$ such that $g(z) = \sum_{j = 0}^\infty  y_{j+1}(z-z_0)^j A(z)^{-1}$ is analytic at $z_0$ and has a zero at $z_0$. We will say the order of the pole of $A(z)$ with eigenvector $y_1$ at $z_0$ is the order of $g(z)$ at $z_0$.
\end{definition}

\begin{remark}
{Definition \ref{def:eig_vectors} is most intuitive in the case where zeros and poles do not overlap and have order and multiplicity one.  In this case, the auxiliary vectors $(x_2, x_3, \ldots)$ and $(y_2, y_3, \ldots)$ are unnecessary, and an eigenvector $x_1$ for zero(I) $z_0$ occurs when $f(z) = A(z) x_1$ is zero-valued at $z_0$, and a pole vector $y_1$ for pole(I) $z_0$ occurs when $g(z) = y_1 A(z)^{-1}$ is zero-valued at $z_0$.}
\end{remark}

\section{Finite-Dimensional Linear Quantum Stochastic Systems (LQSS)}

\label{sec:LQSS}

We provide a brief introduction here for (finite-dimensional) LQSS. Consider a system with a finite-dimensional state $x(t)$ driven by a bosonic input field $b_{\text{in}}(t)$ and generating a bosonic output field $b_\text{out}(t)$. The most general form of a linear system preserving the commutation relations in the input-output relationship has the form
\begin{align}
\frac{\d}{\d t} x &= (A_- x + A_+ x^\#) + (B_- b_{\text{in}} + B_+ b_\text{in}^\#) \\
b_{\text{out}} &= (C_- x + C_+ x^\#) + (D_- b_\text{in}  + D_+ b_\text{in}^\#).
\end{align}
Using the notation $\breve M = \Delta (M_-, M_+)$ for matrices and $\breve v = (v, v^\#)^T$ for vectors {(e.g. as used in~\cite{gough2009enhancement})}, this can be written as
\begin{align}
\frac{\d}{\d t}  \breve x &= \breve A \breve x + \breve B \breve b_\text{in} \\
\breve b_{\text {out}} &= \breve C \breve x + \breve D \breve b_{\text{in}}.
\end{align}
The matrices $\breve A, \breve B, \breve C, \breve D$ must satisfy special physical realizability constraints (e.g.~\bibref{shaiju2012frequency}). 

 The input-output relationship may be characterized by the transfer function of the system. This function is obtained by taking the Laplace transform of the equations of motion, defined by $Y(z) = \int_0^\infty e^{-zt} y(t) dt$. Letting $X(z), B_\text{in}(z)$, and $B_\text{out} (z)$ be the Laplace transforms of $x(t), b_\text{in}(t)$, and $b_\text{out}(t)$, respectively, and eliminating $X(z)$ from the resulting equations, one obtains $B_\text{out}(z) = T(z) B_\text{in}(z)$, where $T(z) = \breve D + \breve C (zI - \breve A)^{-1} \breve B$. In Section~\ref{sec:delayed_feedback} we will build upon finite-dimensional LQSS, introducing time-delayed feedback. 
 %\red{One-sentence remark on relation to Fourier domain transfer function.}
We will do so by considering the Laplace transform of the newly constructed network including the time-delayed feedback.
 
 Theorem 4 in~\bibref{shaiju2012frequency} states an equivalent condition for  physical realizability in the frequency domain when $\breve A, \breve B, \breve C, \breve D$ are doubled up in the sense of \eqref{eq:dblmat} and assuming the state space realization is minimal and the eigenvalues of $\breve A$ satisfy $\lambda_i (\breve A) + \lambda_j (\breve A) \ne 0$ {(a condition which always holds for stable systems with $A < 0$)}. With these assumptions the system is physically realizable if and only if
\begin{enumerate}
\item The transfer function $T(z)$ is $J$-unitary.
\item The matrix $\breve D$ has form $\breve D = \Delta (S, 0)$ for a unitary matrix $S$.
\end{enumerate}
The first condition ensures the commutation relations are preserved, while the second condition ensures there is no static {(infinite-bandwidth)} squeezing in the system {(this would lead to energy divergences and breaks the Hudson-Parthasarathy quantum stochastic calculus)}.  We will show  in the following sections that the state space matrices being doubled-up and the first condition above can be generalized for the cases when there is no finite-dimensional state-space. However, the second condition has no clear analogue (see in particular the discussion in Section~\ref{sec:static_squeezing}).

\section{Linear Quantum Stochastic Systems with Delayed Feedback}
\label{sec:delayed_feedback}

{We begin with an LQSS as described in Section~\ref{sec:LQSS} having $2(N+M)$ inputs and $2(N+M)$ outputs. $N+M$ of the input/output ports correspond to the input/output annihilation field, and the other $N+M$ of the input/output ports correspond to the input/output creation field.
 For convenience, we permute the input and output ports, so that the odd-labeled ports correspond to the creation fields and the even ports to the creation fields.}
 
 We assume this transfer function corresponds to a finite-dimensional (i.e. it is a proper rational function) throughout (see Assumption~\ref{assumption:finite_dim}).
%Label the first $2N$ input and output channels $\mathbf{x}_\text{in}$ and $\mathbf{x}_\text{out}$, respectively. \red{(Labels never used in paper)} 
%Similarly, label  the {last} $2M$ input and output channels $\mathbf{y}_\text{in}$ and $\mathbf{y}_\text{out}$, respectively. 
The transfer function of this system has form
\begin{align}
T (z) = 
\begin{pmatrix}
T_1(z) && T_2(z) \\
T_3(z) && T_4(z)
\end{pmatrix},
\end{align}
where the matrix has been partitioned into blocks of size $2N\times2N$, $2N\times2M$, $2M\times2N$,  and $2M\times2M$.

%\red{I think that a figure showing a dummy system (just a box with inputs / outputs and a delayed feedback) would be helpful here.}

\emph{Remark:} A complex-valued function on the extended plane is a proper rational function if and only if  it is meromorphic on the extended complex plane and analytic at infinity. This applies to the components of $T(z)$, and will be important throughout.

\begin{figure}[h]
\includegraphics[width=0.5\textwidth]{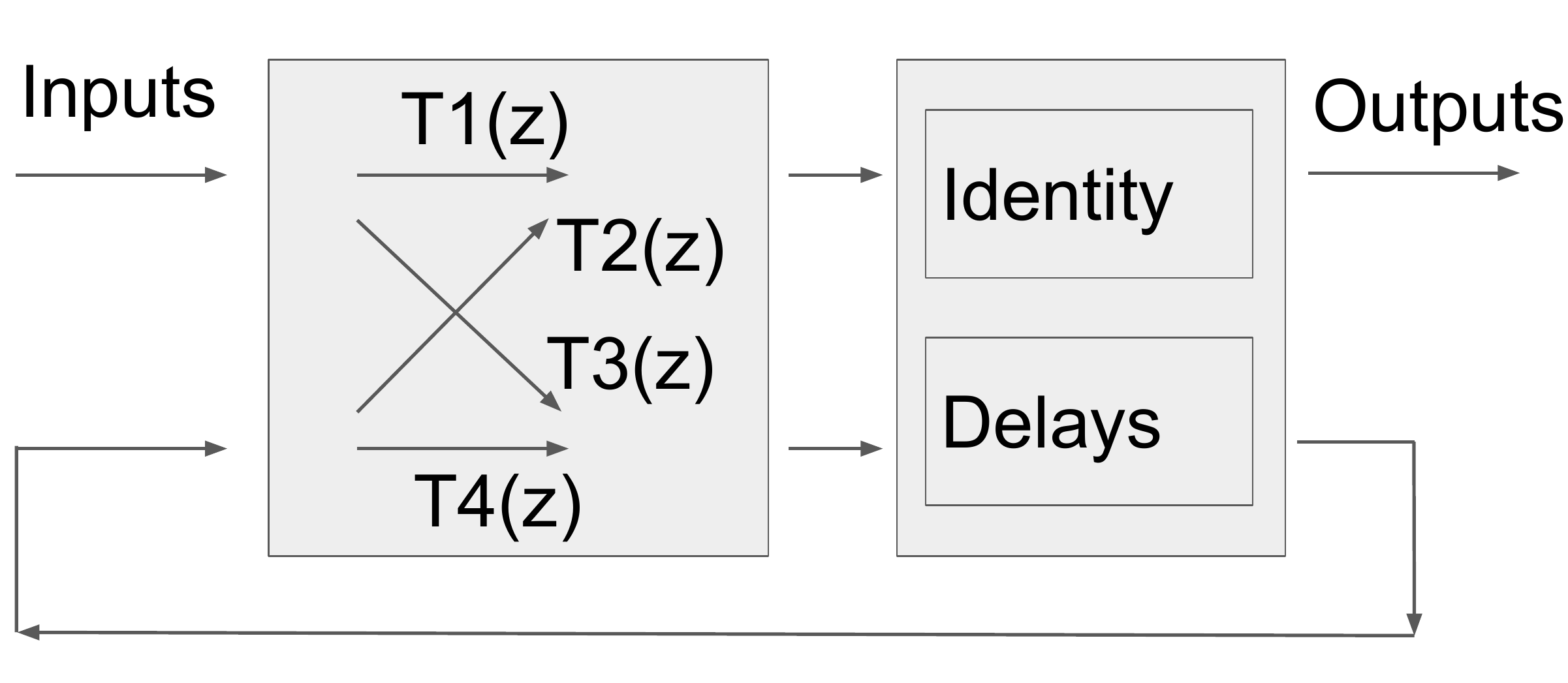}
\caption{{The setup of the network with delayed feedback constructed in Eq.~(\ref{equ:tilde_T_construction}). The arrows in the left box indicate how the blocks $T_i(z)$ for $i=1,2,3,4$ relate the inputs and outputs of $T(z)$ to the inputs and outputs of $\tilde T(z)$ with respect to the time delays.}}
\label{fig:input-output}
\end{figure}

Next, we add delays with feedback to the system.
In the case for linear quantum systems, we use
\begin{align}
\label{equ:tilde_T_construction}
E(z) &= \text{diag}(e^{-T_1 z},e^{-T_1 z},...,e^{-T_M z},e^{-T_M z}). \\
\tilde T(z) &= 
T_1(z) +T_2(z) E(z)(I - T_4(z) E(z))^{-1} T_3(z) \\
 &= 
T_1(z) +T_2(z) (E(-z) - T_4(z))^{-1} T_3(z).
\end{align}
Each time delay occurs in two terms of $E(z)$, once for annihilation and once for the creation field.
Notice that $E(-z) = E(z)^{-1}$ because of its special form.
This notation will be used throughout.
{Figure~\ref{fig:input-output} shows conceptually how the blocks $T_i(z)$ for $i=1,2,3,4$ are used to construct the new transfer function $\tilde T(z)$.}
It can be shown $E(z)$ is $J$-unitary. This is a similar setup to~\cite{gough2010squeezing}.
This term can be derived from the Fourier transform on an linear quantum system with delayed input ports.

\subsection{Assumptions}

\label{sec:assumptions}

These are assumptions we will make throughout.

\begin{enumerate}
\item The transfer  function $T(z)$ is a proper rational function. 
This means that each of its components can be expressed as a ratio of polynomials, and that the degree of the numerator is never greater than the denominator.
 Notice in particular this implies $T(z)$ is analytic at infinity.
 \label{assumption:finite_dim}
\item The transfer function $T(z)$ is doubled-up and $J$-unitary.
\label{assumption:physically_realizable}
\item There is some $C$ so that for $\{ z: |z| > C \}$, $T_2(z), T_3(z),T_4(z)$ are invertible, and their singular values are bounded away from zero by some positive constant. This generalizes the condition in section 8.1 of reference~\cite{tabak2016trapped} where the network components were static and passive. Here, $T_4(z)$ replaces the static matrix $M_1$.
{This condition on $T_4(z)$ excludes a series product with delay $B \triangleleft A$. For systems that do not satisfy this assumption, an approach similar to that used in~\cite{tabak2016trapped} may be used in order to separate out terms corresponding to feedforward time delays. The condition on $T_2(z)$ and $T_3(z)$ allows us to prove Lemma~\ref{lemma:poles_f}.}
%\red{(This excludes a series product with delay $B \triangleleft A$, which could also be interesting, but not treatable here (I think the $B(z)$ ends up being non-constant).  But I think that the Grimsmo paper treats the series-product case, so you can cite it to differentiate your work.)}
\label{assumption:proper}
%\item  For simplicity, that the partial multiplicities of the zeros and poles of $\tilde T$ are first-order, in the sense discussed in~\cite{prather1987factorization}.
\item  periods $T_1,...,T_N$ are commensurate. This suffices for practical purposes, and simplifies some of the proofs.
\label{assumption:commensurate}
\item Zeros(I) and poles(I) from Definition~\ref{def:zeros_I} do not occur in the same location. \label{assumption:zeros}
\item For simplicity,  suppose that whenever we find a zero or pole of a transfer function of interest, there is up to a scalar only one eigenvector at the zero/pole (i.e. the multiplicity of each zero/pole is one). Also, each zero/pole has order one in the sense of Definition~\ref{def:eig_vectors}.
\label{assumption:simple}
\end{enumerate}

%FUTURE WORK: Following~\cite{tabak2016trapped}, we will show how a component consisting only of feedforward delays and passive scattering components can be separated out from the cascade in the case $T_4(z)$ is not invertible, in such  a way that the remaining system has a different transfer function with a $T_4(z)$ that is invertible. This will rely on the fact that for linear systems, parallel delays commute through components. The approach will be similar to~\cite{tabak2016trapped}, section 8.

There are additional specific non-degeneracy assumptions discussed in Section~\ref{sec:construction_of_Ps}.

\section{Properties of transfer functions}

\label{sec:TF_claims}

\begin{claim}
\
\begin{itemize}
\label{claim:J_properties}
\item If two transfer functions are $J$-unitary, then so is their product.
\item If two transfer functions can be written in doubled-up notation, then so can their product.
\item The inverse of a $J$-unitary function is $J$-unitary.
\end{itemize}
\end{claim}

\emph{proofs:} Direct computation.

\begin{claim}
\label{claim:inverse_doubled_up}
If a rational transfer function is doubled-up, then so is its inverse.
\end{claim}

\emph{proof:} This follows using a realization of the inverse of a transfer function, which is valid {when} $D$ is invertible (see e.g. reference~\cite{Lall2008}):
% involving the Shur complement (see e.g.~\cite{horn2012matrix}), applied to the ABCD matrices yielding the transfer function:

%
%\begin{align}
%\begin{pmatrix}
%A && B \\
%C && D
%\end{pmatrix}^{-1}
%=
%\begin{pmatrix}
%(A-BD^{-1}C)^{-1} && -A^{-1} B (D - C A^{-1} B)^{-1} \\
%-(D - C A^{-1} B)^{-1} C A^{-1} && (D - CA^{-1} B)^{-1}
%\end{pmatrix}
%\end{align}
%\begin{align}
%\begin{array}[c|c]
%A & B\\
%  \hline
%C & D
%\end{array}
%\end{align}

% \begin{align}
% 	\left(
% 	\begin{array}{c|c}
% 	A & B \\
% 	\hline
% 	C & D\\	
% 	\end{array}
% 	\right)^{-1}
% 	=
% 	\left(
% 	\begin{array}{c|c}
% 	K^{-1} & -K^{-1}B D^{-1} \\
% 	\hline
% 	-D^{-1}C K^{-1} & D^{-1}(C K^{-1} B D^{-1})
% 	\end{array}
% 	\right).
% \end{align}
% where $K \equiv A - B D^{-1} C$.

Denoting a transfer function by $G(z) = D + C(zI - A)^{-1}B$ 
by $\left(
	\begin{array}{c|c}
	A & B \\
	\hline
	C & D\\	
	\end{array}
	\right) (z)$,
its inverse is

\begin{align}
	\left(
	\begin{array}{c|c}
	A & B \\
	\hline
	C & D\\	
	\end{array}
	\right)^{-1} (z)
	=
	\left(
	\begin{array}{c|c}
	A - B D^{-1} C & -BD^{-1} \\
	\hline
	D^{-1} C & D^{-1}
	\end{array}
	\right) (z).
\end{align}
From this one sees that the matrices 
for the state space representation 
of the inverse are also doubled up.
$\blacksquare$

\begin{claim}
$J$-unitarity is preserved under concatenation, series, and feedback products.
\end{claim}

\emph{proof:} The case for the concatenation product is trivial:
\begin{equation}
	\begin{bmatrix} T_1(z) & 0 \\ 0 & T_2(z) \end{bmatrix} J \begin{bmatrix} T_1(z) & 0 \\ 0 & T_2(z) \end{bmatrix}^\dagger
	= \begin{bmatrix} T_1(z) J_1 T_1(z)^\dagger & 0 \\ 0 & T_2(z) J_2 T_2(z)^\dagger \end{bmatrix} = \begin{bmatrix} J_1 & 0 \\ 0 & J_2 \end{bmatrix} = J
\end{equation}
The series product follows from associativity:
\begin{equation}
	\bigl(T_1(z)T_2(z)\bigr) J \bigl(T_1(z)T_2(z)\bigr)^\dagger
	= T_1(z) \bigl(T_2(z) J T_2(z)^\dagger\bigr) T_1(z)^\dagger = T_1(z) J T_1(z)^\dagger = J
\end{equation}
The only nontrivial case is the feedback product.  Here we start with a transfer function
\begin{equation}
	T(z) = \begin{bmatrix} T_1(z) & T_2(z) \\ T_3(z) & T_4(z) \end{bmatrix}
\end{equation}
$J$-unitarity implies:
\begin{align}
	T_1 J T_1^\dagger + T_2 J T_2^\dagger & = J \nonumber \\
	T_1 J T_3^\dagger + T_2 J T_4^\dagger & = 0 \nonumber \\
	T_3 J T_1^\dagger + T_4 J T_2^\dagger & = 0 \nonumber \\
	T_3 J T_3^\dagger + T_4 J T_4^\dagger & = J \label{eq:t1234subs}
\end{align}
Expanding $T J T^\dagger$ using $T = T_1 + T_2(1 - T_4)^{-1} T_3$, and using the rules (\ref{eq:t1234subs}) to trade $T_1, T_3$ for $T_2, T_4$, we find:
\begin{align}
	T(z) J T(z)^\dagger & = J + T_2(1 - T_4)^{-1} \Bigl[-(1 - T_4)J(1 - T_4^\dagger) - T_4 J (1 - T_4^\dagger) \nonumber \\
	& \qquad\qquad\qquad\qquad\quad - (1 - T_4) J T_4^\dagger - T_4 J T_4^\dagger + J\Bigr] (1 - T_4^\dagger)^{-1} T_2^\dagger
\end{align}
The terms inside the square brackets vanish, leaving $J$, so $T(z)$ is $J$-unitary. $\blacksquare$

\begin{claim}
If $T(z)$ is unitary ($J$-unitary), then so is $\tilde T(z)$ defined above.
\end{claim}

\emph{proof:} $\tilde{T}(z)$ is be obtained from $T(z)$ and $E(z)$ using concatenation, series and feedback operations.  Since these preserve $J$-unitarity, $\tilde{T}(z)$ must be $J$-unitary as well. $\blacksquare$

%Using
%\begin{align}
%T(z) J = J T(z)^\dag,
%\end{align}
%we find
%\begin{align}
%T_1(z) J = J T_1 ^\dag, T_2(z)J = J T_3(z)^\dag; T_3(z)J = J T_2(z)^\dag; T_4(z) J = J T_4(z)^\dag.
%\end{align}
%Using these relations,
%
%\begin{align}
%&(T_1(z) +T_2(z) E(z)(I - T_4(z) E(z))^{-1} T_3(z)) J 
%\\=& J (T_1(z) + T_3(z)^\dag (I - E(z)^\dag T_4(z)^\dag)^{-1} E(z)^\dag T_2(z)^\dag.
%\end{align}

\subsection{Properties of the Roots and Poles of $\tilde T (z)$.} 

\hfill 

\begin{claim}
\label{claim:vectors_zeros_poles}
If a transfer function $T(z)$ is $J$-unitary, its (complex) poles and zeros come in pairs of the form $z_0,-\overline{z}_0$. Furthermore, suppose $x$ is an eigenvector of the zero $z_0$ of $T(z)$ in the sense of~\cite{prather1987factorization}. Then $x^\dag J$ is an eigenvector of the pole $-\overline{z}_0$ at $T(z)$ in the sense of~\cite{prather1987factorization}.
\end{claim}

\emph{proof:}
This follows from $ T(z) J T(- \overline z)^\dag = J$.
In particular, we can re-arrange to get  
\begin{align}
\label{equ:arranged_J_unitary}
T^{-1} ( z) = J T^\dag (- \overline z) J.
\end{align}
Since $z_0$ is an eigenvalue of $T(z)$ with eigenvector $x$, there are some vectors $x_2,x_3,...$ such that
{$T(z) [ x + x_2(z-z_0) + ...]$} is analytic with a zero at $z_0$. Multiplying on the left of Eq.~(\ref{equ:arranged_J_unitary}) by {$x^\dag J +x_2^\dag J (z + \overline {z_0}) + ...$}, we find the expression is analytic with a zero at $-\overline z_0.$ 
$\blacksquare$

\begin{claim}
\label{claim:doubled_up}
 If a rational transfer function $T(z)$ is doubled-up, then it satisfies
\begin{align}
T(z) = \Sigma T(\overline z)^{\#} \Sigma.
\end{align}
\end{claim}

\emph{proof:} 
$T(z)$ is doubled-up if and only if for each matrix $R$ among $A,B,C,D$, $\Sigma R \Sigma = R^\#$. The rest is straightforward algebra.
$\blacksquare$

%For the converse direction, notice that if $ABCD$ is a minimal realization, then it is algebraically equivalent to any other minimal realization $\bar A \bar  B \bar C\bar D$~(Theorem 7.M3 in \cite{chen1995linear}) . Explicitly, this means that for some invertible matrix $P$,
%\begin{align}
%\bar A = PAP^{-1} && \bar B = PB && \bar C = CP^{-1} && \bar D = D,
%\end{align}
%where the internal degrees of freedom of $\bar A \bar  B \bar C\bar D$ satisfy $\bar x(t) = P x(t)$. But if the internal degrees are fixed, then $P = I$, so the $ABCD$ model is unique. This proves the converse.

The property $T(z) = \Sigma T(\overline z)^{\#} \Sigma$ generalizes {the} notion of a doubled-up transfer function for a finite dimensional system to more general functions (i.e. $T(z)$ is no longer required to be rational).

%    \emph{claim:} Given a doubled-up rational transfer function $T(z)$, if $z$ is a pole of $T(z)$, then so is $\overline z$.
%    
%    \emph{proof:}
%    Notice $A= \Sigma A^\# \Sigma,$ and therefore
%    \begin{align}
%    \det (Iz - A) = 0
%    \implies
%    \det(I \overline z - A)
%    = \det (I\overline z - A^\#) 
%    = 0.
%    \end{align} 

\begin{claim}
\label{claim:doubled_zeros}
Given a transfer function $T(z)$ satisfying the property $T(z) = \Sigma T(\overline z)^{\#} \Sigma$, if $z_0$ is a zero of $T(z)$, then so is $\overline {z_0}$. If $x$ is an eigenvector of $T(z)$ at $z_0$, then $\Sigma x^\#$ is an eigenvector of $T(z)$ at $\overline {z_0}$.

 Similarly, if $z_0$ is a pole of $T(z)$, then so is $\overline {z_0}$. If $x$ is an the eigenvector of $T(z)$ at $z_0$, then $\Sigma x^\# $ is an eigenvector of $T(z)$ at $\overline{z_0}$.
 \end{claim}

\emph{proof:} Since $z_0$ is a zero of $T(z)$,  
 there are some vectors $x_2,x_3,...$ such that
{$T(z) [ x + x_2(z-z_0) + ...]$} is analytic with a zero at $z_0$. Applying the property  $T(z) = \Sigma T(\overline z)^{\#} \Sigma$, and using the fact that $f(z)$ is analytic if and only if $\overline{ f(\overline z)}$ is analytic, we obtain the desired result.

The proof for the poles follows similarly. $\blacksquare$

\begin{corollary}
\label{equ:real_eigenvalue_doubled_up}
If the zero $z_0$ with eigenvector $x$ is real, then $y = \Sigma x^\#$ is also an eigenvector at $z_0$.
\end{corollary}

\begin{claim}
 Suppose $T(z)$ is a rational transfer function that can be written in doubled-up form. 
%Assume that $\tilde T(z)$ preserves the poles of $T(z)$ (i.e. if $z$ is a pole of $T(z)$ then it is also a pole of $\tilde T(z)$   ). 
If $z$ is a pole of $\tilde T(z)$, then so is $\overline z$. 
\end{claim}

\emph{proof:} This follows from the fact that $\Sigma \tilde{T}(\overline{z})^{\#} \Sigma = \tilde{T}(z)$, proved using the identities $\Sigma E(z)\Sigma = E(\overline{z})^{\#}$ and $\Sigma A \Sigma = A^{\#}$, etc., below:
\begin{align}
	\Sigma \tilde{T}(\overline{z})^{\#} \Sigma
	& = \Sigma \left[ D^{\#} + C^{\#}(z E(-\overline{z})^{\#} - A^{\#})^{-1} B^{\#}\right] \Sigma \nonumber \\
	& = D + C (z E(-z) - A)^{-1} B = \tilde{T}(z)
\end{align}
From here we use Claim~\ref{claim:doubled_zeros} to show that $\bar{z}$ is also a pole of {$\tilde{T}(z)$}.
$\blacksquare$

\subsection{The poles of $\tilde T(z)$ have bounded real part}

We begin by showing that if $T_4(z) $ has singular values bounded away from zero  for all $z \in \{ |z| > C \}$ for some $C$ (Assumption~\ref{assumption:proper}), 
then the poles of $\tilde T(z)$ have bounded real part.
We will also use that $T_4(z)$ is a rational function (Assumption~\ref{assumption:finite_dim}), and therefore its singular values are also bounded from above.
%then we can find a lower bound for the real part of the poles of $\tilde T(z)$.

% Furthermore,  $T_4(z)$ has singular values bounded from above for sufficiently large $\Re(z)$, since $T(z)$ is a rational function. Using this, we will bound the poles of $\tilde T(z)$ from above as well. 

The original transfer function $T(z)$ is rational, so it has a finite number of poles. Therefore for the purposes of showing a bound for the real part of the poles of $\tilde T(z)$, we can ignore the poles of $T(z)$.
%, and for computing the Nevanlinna order (defined and discussed below), we can neglect them without a loss of generality.
The remaining poles of $\tilde T(z)$ can only occur when $(E(-z) - T_4(z))^{-1}$ has  a pole.
We will obtain our desired bounds using the following lemma:

\begin{lemma}
\label{lemma:singular_values}
Given matrices $A,B$ with singular values satisfying $\sigma_{\min}(A) - \sigma_{\max} (B) > \epsilon$, the matrix $A-B$ is invertible, and $\sigma_{\min} (A-B) > \epsilon$. 
\end{lemma}

\emph{proof:} {The key step employs the triangle inequality:}
\begin{align}
\sigma_{\min}(A-B) &= \inf_{\| x\| = 1} \| (A-B) x \|
\ge \inf_{\| x\| = 1} (\| A x \| -  \| B x \|) \nonumber \\
& \ge \inf_{\|x\| = 1} \| A x \| - \sup_{\|x\| = 1} \| B x \|
 = \sigma_{\min}(A) - \sigma_{\max} (B) > \epsilon.   
\end{align}
The matrix $A-B$ is invertible since all of its singular values are positive.
$\blacksquare$

\begin{lemma}
If $T_4(z) $ has singular values bounded away from zero for all $z \in \{ |z| > C \}$ for some $C$, then there is a strip $\mathcal{S} = \{ z : C_{low} < \Re(z) < C_{high} \}$ outside of which  $\|\tilde T(z)\|$ is bounded from above.
\label{lemma:strip}
\end{lemma}

\emph{proof:}
We assumed that  for  all $z \in \{ |z| > C \}$ for some $C$, the smallest singular value $\sigma_{\min} (T_4(z))$ is bounded away from zero. 
Notice when $\Re(z)<0$, the greatest singular value of $E(-z)$ is $\exp(T_{\min} \Re(z))$, where $T_{\min}$ is the shortest delay. 
Thus when $\Re(z)$ becomes sufficiently negative, the greatest singular value of $E(-z)$ is smaller than the smallest singular value of $T_4(z)$, which is bounded from below. 
 Hence, using Lemma~\ref{lemma:singular_values}, $E(-z) - T_4(z)$ is invertible for some $\Re(z) < C_{low}$, and its smallest singular value is bounded from below. A bound can be computed explicitly in terms of $T_{min}$ and the lower bound on $\sigma (T_4(z))$.
 
  Similarly, as $\Re(z)$ grows in the positive direction, the smallest singular value of of $E(-z)$ diverges. Using the boundedness of the singular values of $T_4(z)$ for sufficiently large $\Re(z)$ and  applying Lemma~\ref{lemma:singular_values},  we find 
  $E(-z) - T_4(z)$ is invertible for some $\Re(z) > C_{high}$, and that its singular value is bounded from below.
  
  Combining the two results above,  we obtain a strip $\mathcal{S} = \{ z : C_{low} < \Re(z) < C_{high} \}$ such that all the poles of $\tilde T(z)$ occur inside $\mathcal{S}$, and the smallest singular value of $E(-z) - T_4(z)$ is bounded from below outside the strip. Thus we can bound $\|(E(-z) - T_4(z))^{-1}\|$ and therefore $\|\tilde T(z)\|$ from above outside $\mathcal{S}$. 
   $\blacksquare$

\section{Approximation using a static component}

\label{section:static}

We will construct a transfer function $\tilde S(z)$, which behaves similarly to $\tilde T(z)$ for large $|z|$ by replacing $T(z)$ with a constant $S$. This corresponds to replacing the internal components of the system with static counterparts. $\tilde S(z)$ is simpler to analyze than $\tilde T(z)$, and will be useful for several of the proofs.

\subsection{Relation Between $\tilde T(z)$ and $\tilde S(z)$ }

Since we have assumed that $T(z)$ is finite-dimensional, in the limit $| z | \to \infty$ we have $T(z) \to S$, where $S$ is a constant $J$-unitary, doubled-up matrix. 
We partition $S$ in the same way as we did $T(z)$ into four blocks of the same sizes:
\begin{align}
S = \lim_{|z|\rightarrow \infty} T(z) = \begin{pmatrix}
S_1 & S_2 \\
S_3 & S_4 
\end{pmatrix}.
\end{align}
Define
\begin{align}
\label{equ:S_tilde}
\tilde S(z) = S_1 + S_2  E(z) \left( I - S_4 E(z) \right)^{-1} S_3.
\end{align}
Finding a convergent factorization for $\tilde S(z)$ in terms of canonical factors will be a precursor to a similar factorization for $\tilde T(z)$. The relationship between $\tilde S$ and $\tilde T(z)$ for large $|z|$ is described in Lemma~\ref{lemma:static_approximation} below.

It will be useful to introduce $f_T(z) = \det (I - T_4(z) E(z))$ and $ f_S(z) = \det (I - S_4 E(z))$. 

\begin{lemma}
\label{lemma:poles_f}
Restricted to sufficiently large $|z|$, the poles of $\tilde T(z)$ and $\tilde S(z)$, respectively,
 are the roots of the functions $f_T(z)$ and $f_S(z)$. 
 \end{lemma}
 
 \emph{proof:}
This follows using Assumption~\ref{assumption:proper} and that each $T_i(z)$ is rational ($i=1,2,3,4$). Specifically, the poles of  $T_i(z)$ are bounded, so when $|z|$  is large enough $\tilde T(z)$ can only have poles when $f_T(z)$ has zeros (and similarly for $\tilde S(z)$). By Assumption~\ref{assumption:proper}, when $|z|$ is large enough, $T_2(z)$ and $T_3(z)$ have singular values bounded away from zero, and hence $S_2$ and $S_3$ have no zero singular values. Therefore, all of the zeros of $f_T(z)$ and $f_S(z)$ for large enough $|z|$ are also poles of $\tilde T(z)$  and $\tilde S (z)$, respectively.
$\blacksquare$

\begin{lemma} Rate of convergence.
\label{lemma:rate_of_convergence}

(i) The rate of convergence of $T_4(z) \to S_4$ is $O(1/|z|)$.

(ii) When $\Re (z)$ is bounded, the rate of convergence of $f_T(z) \to f_S(z)$ is $O(1/|z|)$.
\end{lemma}

\emph{proof (i):}
From Assumption~\ref{assumption:finite_dim}, $T_4(z)$ is analytic at infinity. Since $T_4(z) \to S_4$ as $|z| \to \infty$, the Laurent series of each of the components of $T_4(z)$ at infinity has the form $\sum_{n=0}^\infty \frac{c_n}{z^n}$. This implies the rate of convergence of $T_4(z) \to S_4$ is $O(1/|z|)$. 
$\blacksquare$

\emph{proof (ii):}
Expand the determinant as a polynomial function for $f_T(z)$ and $f_S(z)$. Notice for both $f_T(z)$ and $f_S(z)$, the coefficients and the entries of $E(z)$ are the same. The desired estimate follows by subtracting the two expressions, using the triangle inequality, and noting each of the entries of $E(z)$ is bounded when $\Re(z)$ is bounded.
$\blacksquare$

\begin{lemma}
\label{lemma:static_approximation}
When $|z| $ is large,  $\tilde T(z)$ approaches $\tilde S(z)$ in the following sense. For sufficiently large $M$, there exists $\epsilon_M = O(M^{-1})$ such that the following hold:

%Suppose we are given $\epsilon > 0$. Then there exists $M >0$ such that for $|z| > M$, 
%(i) $z$ is inside a ball of radius $\epsilon$ of a zero (pole) $z_0$ of $\tilde T(z)$ if it is also inside a ball of radius $O(\epsilon)$ of a zero (pole) $w_0$ of $\tilde S(z)$ (and vice versa). 
(i)  For each zero (pole) $w_0$ of $\tilde S(z)$ satisfying $|w_0| > M$, there is exactly one zero (pole) $z_0$ of $ \tilde T(z)$ inside $B_{\epsilon_M} (w_0)$, and $w_0$ is the only zero (pole) of $\tilde{S}(z)$ in $B_{\epsilon_M} (z_0)$. Similarly, the statement holds if we exchange $\tilde S(z)$ and $\tilde T(z)$.

(ii) For $z_0$ and $w_0$ above, the orthogonal projectors $P_T, P_S$ onto the span of the eigenvectors of $\tilde T(z)$ at $z_0$ and $\tilde S(z)$ at $w_0$, respectively, satisfy $\| P_T - P_S \| = O (\epsilon_M)$.

(iii)
Given arbitrary but fixed $\epsilon$,
if $|z|>M$
and $z \notin B_{\epsilon} (w_0), B_{\epsilon} (z_0)$ for all poles 
$w_0$ and $z_0$ of $\tilde S(z)$ and $\tilde T(z)$, then we can estimate $\|\tilde T(z) - \tilde S(z)\| = O(1/M).$
\end{lemma}
 
\emph{The proof is given in Appendix~\ref{appendix:approx}.}

\section{Fundamental Factors of Quantum Linear Systems}

\label{sec:fundamental_factors}

In this section we will use the notation from Definitions~\ref{def:flat} and~\ref{def:Delta}.

Our goal will be to  show that $\tilde T(z)$ can be factorized as $B \prod P_i(z)$ where $P_i(z)$ are elementary transfer functions of physically realizable components and $B$ is a constant $J$-unitary and doubled-up matrix. % \blue{of the form $B = \Delta(S, 0)$ (i.e. $B$ does not include static squeezing terms)}, which does. 
For the $P_i(z)$ terms, we will use the transfer function of a generic physically realizable linear system with only two root and pole pairs. Such a system can be described using the SLH framework (see e.g.~\cite{combes2017slh}) by
\begin{align}
\label{equ:SLH}
(S,L,H) = \left( I_{2\times2}, \Lambda \breve{a}, \frac{1}{2} \breve{a}^* {\Omega} \breve{a} \right).
\end{align}
Above in Eq.~(\ref{equ:SLH}), $\breve a = [a, a^*]^T$, $\Lambda = [\Lambda_-, \Lambda_+]$, and {$ {\Omega} = \Delta(\omega, \epsilon)$} is a Hermitian matrix.  {Specifically, $\breve{a}$ represents the internal field (creation / annihilation operators), $\Lambda$ is related to the Lindblad terms in the master equation (coupling to environment) and $\Omega$ is related to the system Hamiltonian.}
%\red{(I suggest renaming $\Delta \rightarrow \omega$ to avoid the confusing notation $\Delta(\Delta, \epsilon)$.  Note that my notes on canonical form used $\Delta$, so this will have to be changed.)}
Defining $\breve \Lambda = \Delta (\Lambda_-, \Lambda_+)$, the state-space realization of this system is given by
\begin{align}
A &= -\frac{1}{2} \breve \Lambda ^\flat \breve \Lambda - i \Omega J, & B &= -\breve \Lambda^\flat, \\
C &= \breve \Lambda, & D &= I.
\end{align}
We can normalize $\breve \Lambda$ using $V = \breve \Lambda / \sqrt{\kappa}$ with $\kappa = |\Lambda_-|^2 - |\Lambda _+|^2$. This results in the form
\begin{align}
\label{equ:OPO}
P(z) = I - V V^\flat +
V
\frac
{z  - \frac{1}{2}\kappa + i \Omega J}
{z  + \frac{1}{2}\kappa + i \Omega J}
V^\flat.
\end{align}
In Eq.~(\ref{equ:OPO}) above, notice $V$ is a doubled-up matrix satisfying $V^\flat V= I$, and $P_v = V V^\flat$ is the projector onto the subspace in which the input-output field interacts with the system.

\subsection {Canonical form of the fundamental factors }
\label{sec:canonical_opo}
There are two possibilities for the form of the transfer function, depending on whether its roots are complex or purely real. 
%To obtain these canonical forms, we  diagonalize $i\Omega J$. 
%\red{TODO: add details for derivation of canonical form}
If the roots of the transfer function are complex {($|\omega| > |\epsilon|$), they come in pairs $(z_0, \overline{z_0})$, and} the transfer function can be rewritten as:
 
\begin{align}
\label{equ:OPO_complex}
P(z)
 = I - V V^\flat
 + V \begin{pmatrix}
 \frac{z - z_0}{z + \overline z_0} && 0 \\
 0 &&  \frac{z - \overline {z_0}}{z +  z_0}
 \end{pmatrix}
  V^\flat.
\end{align}

If the roots {$(z_1, z_2)$} are purely real {($|\omega| < |\epsilon|$)}, the transfer function can be rewritten as:
\begin{align}
\label{equ:OPO_real}
P(z)
 = I - V V^\flat
 +\frac{1}{2} V 
  \begin{pmatrix}
 1 && i \\
 1 && -i
 \end{pmatrix}
 \begin{pmatrix}
 \frac{z - z_1}{z + z_2} && 0 \\
 0 &&  \frac{z - z_2}{z +  z_1}
 \end{pmatrix}
  \begin{pmatrix}
 1 && 1 \\
-i && i
 \end{pmatrix}
  V^\flat.
\end{align}

To derive this form, first write the form of the Hermitian matrix $\Omega = \begin{pmatrix} a & b\, e^{i\phi} \\b\, e^{-i\phi}  & a
\end{pmatrix}$, where $a, b > 0$ without loss of generality (the $a < 0$ case is straightforward to transform into this form). 

{In the case when 
$i \Omega J$ has complex roots, we find $a > b$.  Using $\eta = \tanh^{-1}(b/a)$ and} {$c = \sqrt{a^2 - b^2}$,} {we can write
$\Omega = c \begin{pmatrix} \cosh(\eta) & \sinh(\eta)e^{i\phi} \\ \sinh(\eta)e^{-i\phi} & \cosh(\eta)
\end{pmatrix} = c \begin{pmatrix} \cosh(\eta/2) & \sinh(\eta/2)e^{i\phi} \\ \sinh(\eta/2)e^{-i\phi} & \cosh(\eta/2)
\end{pmatrix}^2 = cS^2
$. Since $S$ is unitary and $J$-unitary, $i\Omega J = i c S^2 J = i c SJS^{-1}$. Finally, $S$ is also doubled-up, so it can be absorbed into $V$ in Eq.~(\ref{equ:OPO}). This leads to Eq.~(\ref{equ:OPO_complex}) with $z_0 = -\frac{1}{2} \kappa +i c$.
}

In the case when $i \Omega J$ has real roots, we find $a < b$.  Using $\eta = \tanh^{-1}(a/b)$ and {$c = \sqrt{b^2 - a^2}$,} we can write
$\Omega = c \begin{pmatrix} \sinh(\eta) & \cosh(\eta)e^{i\phi} \\ \cosh(\eta)e^{-i\phi} & \sinh(\eta)
\end{pmatrix}.$
$\Omega$ can then be transformed to its canonical form $\begin{pmatrix}
0 & i c \\ -i c & 0
\end{pmatrix}$ using the Bogoliubov transformation $\Omega \to S \Omega S^\dag$, with $S = \begin{pmatrix}
\cosh (\eta/2) e^{i (\pi/4-\phi/2)} & -\sinh (\eta/2) e^{i (\pi/4+\phi/2)} \\
-\sinh (\eta/2) e^{i (-\pi/4-\phi/2)}  & \cosh (\eta/2) e^{i (-\pi/4+\phi/2)} 
\end{pmatrix}$ as the J-unitary and doubled-up transformation matrix.  Once we have transformed $\Omega$ to its canonical form, we can diagonalize $i \Omega J = \frac{1}{2}S^{-1} \begin{pmatrix} 1 & i \\ 1 & -i \end{pmatrix} \begin{pmatrix} c & 0 \\ 0 & -c \end{pmatrix} \begin{pmatrix} 1 & 1 \\ -i & i \end{pmatrix} S$, from which \eqref{equ:OPO} can be used to derive the form \eqref{equ:OPO_real} with $z_1 = \tfrac{1}{2}\kappa - c$, $z_2 = \tfrac{1}{2}\kappa + c$.

For the real roots case, the new matrices 
$ \begin{pmatrix}
 1 && 1 \\
-i && i
 \end{pmatrix}$ and  
 $
 \begin{pmatrix}
 1 && i \\
 1 && -i
 \end{pmatrix}
 $
are inserted {explicitly instead of being absorbed into $V$} to preserve the properties of $V$.
It can be checked that both transfer functions Eq.~(\ref{equ:OPO_complex}) and Eq.~(\ref{equ:OPO_real}) satisfy $P(z) J P^\dag(-\overline z) = J$ and 
$P(z) = \Sigma P(\overline z)^{\#} \Sigma$.

\subsection{Constructing elementary factors to match a transfer function at a zero or pole}

\label{sec:construction_of_Ps}

In this section we discuss how to construct transfer functions of the forms in Eq.~(\ref{equ:OPO_complex}) and  Eq.~(\ref{equ:OPO_real})  that match the zeros/poles and eigenvalues of a $J$-unitary and doubled-up transfer function $F(z)$.
We also discuss the conditions under which it is not possible to construct terms of the desired form.

In Section~\ref{sec:zero_pole_matching}, we will show how a particular term $P(z)$ with matching zeros/poles and eigenvalues can be detached from $F(z)$. 
In Section~\ref{sec:fac_QLS} we will discuss how such terms can be detached sequentially, giving the desired factorization.
 
 For the case of a complex root $z_0$ and its eigenvector $v_0$, we use the term in Eq.~(\ref{equ:OPO_complex}) for $P(z)$. 
 In this case another zero $\overline{z_0}$ and its eigenvector $\Sigma v_0^\#$, along with two poles $-\overline{z_0}$ and $-z_0$ and their eigenvectors, $v_0^\dag J$ and $\Sigma v_0^T J$, respectively, are all determined for both $F(z)$ by Claims~\ref{claim:vectors_zeros_poles} and~\ref{claim:doubled_zeros}. 
 \begin{theorem} 
 \label{theorem:complex_case}
 Given a rational $J$-unitary and doubled-up transfer function $F(z)$ with a conjugate pair of complex roots $z_0$ and $\overline{z_0}$ with respective eigenvalues $v_0$ and $w_0$ with $w_0 = \Sigma v_0^\#$ and $v_0^\dag J v_0 \neq 0$, there exists a term of the form Eq.~(\ref{equ:OPO_complex}) with the same zeros and eigenvalues.
 \end{theorem}
 \emph{proof:}
 The proof is constructive:
\begin{enumerate}
\item Normalize $v_0$ so that $v_0^\dag J v_0 = 1$. Since $w_0 = \Sigma v_0^\#$, 	it follows $w_0^\dag J v_0 = 0$. From this it follows $w_0^\dag J w_0 = -1$. 
\item Set $V = [v_0, w_0]$ in Eq.~(\ref{equ:OPO_complex}). From the condition  $w_0 = \Sigma v_0^\#$ it follows $V$ is in doubled-up form. Using the identities in the previous step, one can show $v_0$ and $w_0$ are eigenvectors of $T(z)$ in the sense:
\begin{align}
T(z) v_0 = \frac{z - z_0}{z + \overline{z}_0} v_0, && T(z) w_0 = \frac{z - \overline{z}_0}{z + {z}_0} w_0.
\end{align}
The eigenvalues are analytic in $z$ and have a zeros at $z_0$ and $\overline z_0$, so $v_0$ and $w_0$ are their respective eigenvectors in the sense of Definition~\ref{def:eig_vectors}.
\end{enumerate}
 
 Since the term Eq.~(\ref{equ:OPO_complex}) is also $J$-unitary and doubled-up, the term found by Theorem~\ref{theorem:complex_case} also has the same poles $-\overline{z_0}$ and $-z_0$ and their eigenvectors (up to a scalar), $v_0^\dag J$ and $\Sigma v_0^T J$.
 
 Next, we discuss the case of real roots.
We will assume that the number of real roots is even, so that they can be paired to form terms of the form Eq.~(\ref{equ:OPO_real}). 
Suppose a pair of zeros $z_1$ and $z_2$ of $F(z)$ is given, along with their corresponding eigenvectors $v_1$ and $v_2$. In this case the poles $-z_1$ and $-z_2$ and their eigenvectors $v_1^\dag  J$ and $v_2^\dag  J$ are determined by Claims~\ref{claim:vectors_zeros_poles}.
 \begin{theorem} 
  \label{theorem:real_case}
 Given a rational $J$-unitary doubled-up transfer function $F(z)$ with two real roots $z_1$ and $z_2$ and corresponding eigenvectors $v_1$ and $v_2$ respectively, such that $v_1^\dag J v_2 \neq 0$, there exists a term $P(z)$ of the form Eq.~(\ref{equ:OPO_real}) with the same zeros and eigenvectors.
 \end{theorem}
\emph{proof:}
\begin{enumerate}
\item Using Corollary~\ref{equ:real_eigenvalue_doubled_up} and the assumption that each eigenspace is one dimensional (Assumption~\ref{assumption:simple}), the phase of the eigenvectors can be chosen such that $v_1  = \Sigma v_1^\#$ and $v_2  = \Sigma v_2^\#$. This implies $v_1^\dag J v_1 = v_2^\dag J v_2 = 0$.
\item
There are two remaining degrees of freedom for how the eigenvectors can be chosen, corresponding to their norms. Notice the value $v_1^\dag J v_2$ is always imaginary due to the previous step. We make one of the constraints $v_1^\dag J v_2 = i/2$ (we may have to switch the order to get the right sign). The other constraint, corresponding to the relative norms of the two eigenvectors, can be chosen arbitrarily.
\item Set $V$ by
\begin{align}
V = (v_1, v_2) \begin{pmatrix} 1 & 1 \\ -i & i \end{pmatrix}.
\end{align}
{As a result of of $v_{i} = \Sigma v_{i}^\#$ for $i=1,2$, the  matrix $V$ is doubled-up in the sense of \eqref{eq:dblmat}.}  
%\red{Note that the formula $\Delta(v_1-i v_2, v_1 + i v_2)$ doesn't work because the dimensions are wrong: $v_{1,2}$ are $2N$-dimensional vectors.  We'd have to introduce half-size vectors $v = (w, w^\#)^T$ and define $V$ in terms of those -- too distracting to the reader.}  Inserting this $V$ into (Eq. (~\ref{equ:OPO_real}), we obtain
\begin{align}
T(z) v_1 = \frac{z - z_1}{z + z_2} v_1,&& T(z) v_2 = \frac{z-z_2}{z + z_1} v_2.
\end{align}
The eigenvalues are analytic in $z$ and have a zeros at $z_1$ and $z_2$, so $v_1$ and {$v_2$} are their respective eigenvectors in the sense of Definition~\ref{def:eig_vectors}.
\end{enumerate}
 Since the term  Eq.~(\ref{equ:OPO_real})  is also $J$-unitary and doubled-up, the term found by Theorem~\ref{theorem:real_case} also has he same poles $-z_1$ and $-z_2$ as $F(z)$ as well as the same eigenvalues at those poles.

%
% For the term in (Eq. (~\ref{equ:one_real_root})  we will require a real root $z_0$ and its eigenvector $v_0$.
% In this case the corresponding pole $-z_0$ and its eigenvector $v_0^\dag J $ are determined 
% by claim~\ref{claim:vectors_zeros_poles}.
%[ NOTE: In Ryan's note, the form used for real roots was (Eq. (~\ref{equ:OPO_real}) instead of the form (Eq. (~\ref{equ:one_real_root}) used here. We use the form here because in general we did not assume the transfer function has an odd number of real roots.  In theorem 1 of the note, a condition on the eigenvectors $v_1$ and $v_2$ at two real zeros is given to be able to construct the desired term. The condition is specifically $v_1^\dag J v_2 \ne 0$. ]
% 

\label{sec:special_cases}

\subsection{Another form of the elementary factors}
\label{sec:another_form}

The elementary factors in Section~\ref{sec:canonical_opo} can be re-written in another form, which can be useful.
We will focus here on the terms resulting from a canonical factor with two complex roots, Eq.~(\ref{equ:OPO_complex}). 
We can write this function as
\begin{align}
P(z) = I - V V^\flat + V F(z) V^\flat.
\end{align}
Here $V^\flat V = I$ and $F(z) =\text{diag} \left( a(z), a(\overline z) ^* \right) =  \text{diag}\left(\frac{z - z_0}{z + \overline{z_0}},\frac{z - \overline{z_0}}{z + {z_0}} \right)$. With the substitution $W = VJ$, we obtain
\begin{align}
\label{equ:W_form}
P(z) &= I -  W J W^\dag J + W  J F(z) W^\dag J  
= (J -  W J W^\dag   -  W  J F(z) W^\dag ) J .
\end{align}
Also, $W$ satisfies $W^\dag J W = J$ {since $V^\flat V = I$}.
% $W$ has the form $W = [w_{+,1}, w_{-,1}]$ where the $w_{\pm,1}$ are column vectors. 
% We will next show one can find a $J$-unitary matrix $M$ for which two of the columns are $w_{+,1}$ and $w_{-,1}$.
% Splitting the first $n$ and last $n$ coordinates $w_{\pm,1} = (u_{1,\pm},u_{2,\pm})^T$,  we find $u_{1,+}^2 \ge 1$ and $u_{2,-}^2 \ge 1$ from  $W^\dag J W = J$.  
% To obtain the desired $J$-unitary matrix $M$, we can find vectors $w_{\pm,i}$ for $i = 2,...,n$ in the eigenspaces of $J$ corresponding of eigenvalues $\pm 1$ using Gram-Schmidt separately in the two subspaces, starting with $w_{\pm,1}$.
%The resulting vectors satisfy $w_{q,i}^\dag J w_{r,j} = \delta_{q,r}\delta_{i,j}$ where $q,r \in \{ +,-\}$ and $i,j \in \{1,...,n\}$.
%We can form $M$ using the columns $w_{+,1},w_{-,1}, w_{+,2},w_{-,2},...,w_{+,n},w_{-,n}.$

Next, we will complete the column space $\mathcal{W}$ of $W$, forming a $J$-unitary matrix $M$. Care must be taken since we use the indefinite inner product given by $[x,y] \equiv x^\dag J y$, as opposed to the standard inner product.
In our situation, $\mathcal{W}$ is nondegenerate, meaning that if $x\in\mathcal{W}$ and $[x,y] = 0 $ for all $y \in \mathcal{W}$ then  $x = 0$.
Therefore, we can apply proposition 2.2.2 in~\cite{gohberg2006indefinite}, which implies the orthogonal companion given by $\mathcal{W}^{[\perp]} = \{x \in \mathbb{C}^n | [x,y] = 0 \text{ for all } y \in \mathcal{W} \}$ is the direct complement of $\mathcal{W}$.
This result can be used to construct an orthonormal basis $[x_i,x_j] = \pm \delta_{i,j}$, as discussed in~\cite{gohberg2006indefinite} following proposition 2.2.2. Further, proposition 2.2.3 in~\cite{gohberg2006indefinite} implies we can pick our basis so that $[x_i,x_j] = J_{ij}$. Stacking the columns $M = [x_1| ... |x_{2n}]$, we find $M$ is a $J$-unitary matrix.

 We can use $M$ to write
\begin{align}
\label{equ:J_unitary_form}
P(z) = M 
\text{diag} 
\left( a(z),  -a(\overline z)^*, 1, -1,...,1,-1
\right)
M^\dag J.
\end{align}
This term is analogous to the Blachke-Potapov factors Eq.~(13) in reference [17], albeit there are some differences due to the indefinite inner product used in the construction.

\subsection{Zero and Pole Matching}
\label{sec:zero_pole_matching}
Suppose we are given a transfer function $A(z)$ satisfying $A(z) J A^\dag(-\overline z) = J$ and 
$A(z) = \Sigma A(\overline z)^{\#} \Sigma$. We wish to factorize it according to its roots and poles.
By the claims in Section~\ref{sec:TF_claims}, we can relate the eigenvector of the zero $z_0$ to that of a pole at $-\overline z_0$. Further, if the zeros are purely complex, there is another zero-pole pair $\overline z_0, -z_0$, and the eigenvectors of all four points can be related.

\begin{lemma}
\label{lemma:detach}
Given two transfer functions $A(z)$ and $P(z)$ both with a pole (zero) at $w_0$ of partial multiplicity $1$ (for simplicity) with the same eigenvectors, and no other poles or zeros at $w_0$. Then the function $A(z) P(z)^{-1}$ is analytic at $w_0$ with no zeros or poles.
\end{lemma}

\emph{Proof:}
Let $Q$ be the orthogonal projector onto the span of the eigenvectors, and $W = Q / w_0$. As in the proof of theorem 2.1 of [1], we can factorize a pole from both $A(z)$ and $P(z)$:
\begin{align}
A(z) P(z)^{-1}
&= A(z) (I - z W)(I - z W)^{-1} P(z)^{-1} \\
&= A(z) (I - z W)[P(z)(I - z W)]^{-1} \nonumber
=\tilde A(z) \tilde P(z)^{-1},
\end{align}
where $\tilde A( z) = A(z) (I - z W)$ and $\tilde P( z) = P(z) (I - z W)$.
The partial pole multiplicities of $\tilde A(z)$ and $\tilde P(z)$ at $w_0$ are all smaller than those of $A(z)$ and $P(z)$, respectively, by $1$. 
{The zero multiplicities of $A(z)$ and $\tilde A(z)$ at $w_0$ are the same (the same for tilde $P(z)$ and $\tilde P(z)$).}
Since we have assumed (for simplicity) that the poles all have partial multiplicity $1$, the functions $\tilde A(z)$ and $\tilde P(z)$ are analytic at $w_0$, which has no poles or zeros.
We conclude that $\tilde A(z) \tilde P(z)^{-1}$ is analytic at $w_0$ and has no zeros or poles at $w_0$.

The proof is similar for the case of a zero instead of a pole. In this case, where $Q$ is the orthogonal projector onto the span of the zero eigenvectors and $W = Q / w_0$, we factorize:
\begin{align}
A(z) P(z)^{-1}
&= A(z) (I - z W)^{-1}(I - z W)P(z)^{-1} \\
&= A(z) (I - z W)^{-1} [P(z)(I - z W)^{-1}]^{-1} \nonumber 
=\tilde A(z) \tilde P(z)^{-1},
\end{align}
where $\tilde A(z) =  (I - z W)^{-1}$ and $\tilde P(z) = P(z)(I - z W)^{-1}$. The rest of the argument is similar to the case where $w_0$ is a pole.
$\blacksquare$

%\emph{Lemma:} $J$-diagonalization. Given a matrix $\Omega$, if $J \Omega$ is diagonalizable, then there exists a factorization
%\begin{align}
%\Omega = U D U^\flat,
%\end{align}
%where $D$ is a diagonal matrix, and $U$ is unitary.
%
%\emph{Proof: }
%Since $\Omega J$ is diagonalizable, we can find a factorization
%\begin{align}
%\Omega J = U C U^\dag,
%\end{align}
%for a diagonal matrix $C$ and a unitary matrix $U$.
%Let $D = CJ$. We find
%\begin{align}
%\Omega = U D J U^\dag J = UDU^\flat.  \hfill \square
%\end{align}
%
%In the case of the transfer function of an OPO (Eq. (~\ref{equ:OPO}), the matrix $\Omega$ is symmetric, so $\Omega J$ is also symmetric and hence diagonalizable. Therefore we can find a $J$-diagonalization for $\Omega$. Once this is done, without a loss of generality we can write
%\begin{align}
%P(z) = I - V V^\flat +
%V
%\begin{pmatrix}
%\frac{z - z_0}{z + \overline z_0} && 0 \\
%0 && \frac{z - \overline z_0}{z + z_0}
%\end{pmatrix}
%V^\flat.
%\end{align}

\section{Factorization for quantum linear systems by physically realizable components}
\label{sec:fac_QLS}

Ultimately, we  are interested in the setup given in Section~\ref{sec:overview}. We wish to detach from $\tilde T(z)$ a (possibly infinite) sequence of transfer functions of physically realizable terms, which have the form of \eqref{equ:OPO_complex} or \eqref{equ:OPO_real}, so that the remaining function has no zeros or poles. Explicitly, 
\begin{align}
\label{equ:TF_factorization}
\tilde T(z)  = \prod_n P_n(z) B(z) ,
\end{align}
where $B(z)$ has no zeros or poles.
In order to obtain the product in \eqref{equ:TF_factorization}, the appropriate terms $P_n(z)$ will be detached sequentially.
We will also show $B(z)$ is a constant given our assumptions.
In order to obtain a factorization of the above form, we will use for $P_n(z)$ terms  of the form  \eqref{equ:OPO_complex} or \eqref{equ:OPO_real}, depending whether two real roots or two complex roots are chosen. This is discussed in Section~\ref{sec:fundamental_factors}. 

%Using the properties in Section~\ref{sec:TF_claims} and Lemma~\ref{lemma:detach},  the terms $P_i(z)$ can be constructed one at a time, by detaching (in the complex case) two zeros and two poles using the factor in (Eq. (~\ref{equ:OPO_complex}) or (in the real case) two real roots and real poles using the factor in (Eq. (~\ref{equ:OPO_real}), so that the remaining function still has the physical realizability properties. 
%Notice there are special cases when $P_i(z)$ cannot be constructed, which are discussed in Section~\ref{sec:fundamental_factors}.

Before we discuss systems with feedback, we will first discuss the factorization applied to systems without feedback described by a transfer function  $T(z)$ in Section~\ref{sec:no_feedback_factorization}. This case is easier since the product in \eqref{equ:TF_factorization} is finite. 

When feedback is present, the product in \eqref{equ:TF_factorization} in general may be infinite. In this case, we must show the product converges for some ordering of the factors $P_n(z)$.
To this end, we  introduced $\tilde S(z)$ discussed in Section~\ref{section:static}.
$\tilde S(z)$  behaves similarly to $\tilde T(z)$ for large $|z|$, and is easier to study. Physically, this corresponds to replacing the components yielding  the function $T(z)$  with static counterparts.  We will construct our desired factorization for $\tilde S(z)$ in Section~\ref{sec:static_factorization}. In Section~\ref{sec:T_factorization} we will discuss the more general factorization of $\tilde T(z)$.

%\emph{Rate of convergence:} 
%Since $T(z)$ is a rational function, $|T(z) - S|$ uniformly converges at a rate $1/M$ for $|z| >M$, as does $\delta = | f_T(z) - f_S(z) |$.
%Notice the size of the $\epsilon$ neighborhood in part (i) above can be shrunk proportionally to $\delta ^p$ near a zero of order $p$.

\subsection{Finite dimensional system with no feedback}

\label{sec:no_feedback_factorization}

In this section, we will discuss the factorization of the finite-dimensional system $T(z)$.
We make the same assumptions stated above for $T(z)$. 

As discussed above, we can choose terms $P_n(z)$ using terms of the 
form \eqref{equ:OPO_complex} or \eqref{equ:OPO_real}, depending on whether each root is real or complex, with the caveat that the conditions discussed in Section~\ref{sec:special_cases} must hold at each step of the factorization. This leads to a factorization of the form $T(z) = \prod_{n=1}^N P_n(z) B(z)$. Here $N$ is a finite number since there are a finite number of zeros and poles for $T(z)$.

In the limit $|z| \to \infty$, we find each of the $P_n(z)$ and the $T(z)$ converges to a constant matrix. Therefore, $B(z)$ is a constant matrix (call it $B$). It also satisfies the doubled-up and $J$-unitary properties.  Further, if $T(z)$ is physically realizable, then each of the finite $P_n(z)$ can be inverted showing $B$ is also physically realizable.

Notice the existence of a realization of a finite-dimensional state-space matrices $A,B,C,D$ is automatically implied by the physical realization of a system with the appropriate transfer function.

\subsection{Static System with Nonzero Time Delays}

\label{sec:static_factorization}

In the special case $T(z) = S$, we find $\tilde T(z) = \tilde S(z)$ for $\tilde S(z)$ in \eqref{equ:S_tilde}.

We will show the following:
\begin{theorem}
\label{theorem:convergence_static}
Assume that each term $P_n(z)$ can be constructed sequentially as discussed in  Section~\ref{sec:construction_of_Ps}.
%Also assume for simplicity that the complex roots $Z_C$ of $\tilde S(z)$ can be partitioned into sets $Z_+$ and $Z_-$ with the same periodicity as $Z_C$ and $Z_+ = \{ \overline z : z \in Z_-\}$.
Then there is a way to pick terms $P_n(z)$ so that the product $\prod_n P_n(z)$ converges uniformly on compact sets, $\tilde S(z) = \prod_n P_n(z) B$, and $B$ is a constant matrix.
\end{theorem}

\emph{The proof is given in Appendix~\ref{appendix:convergence_static}.}

\subsection{Finite Dimensional System with Nonzero Time Delays}

\label{sec:T_factorization}

Next, we generalize to the case where $T(z)$ replaces the static component $S  = \lim_{z \to \infty} T(z)$, to obtain a factorization for $\tilde T(z)$ instead of $\tilde S(z)$. 

As in the case using a static component in Section~\ref{sec:static_factorization}, again we only need to consider terms $P_n(z)$ with complex roots for convergence, since the number of terms with real roots is finite. For this reason we will again ignore the real roots.

\begin{theorem}
\label{theorem:convergence_general}
Suppose that the factorization for $\tilde S(z)$ in Theorem~\ref{theorem:convergence_static} exists and converges uniformly on compact sets. 
Assume that terms $P_n(z)$ can be detached sequentially from $\tilde T (z)$ as discussed in Section~\ref{sec:construction_of_Ps}.
Then there is a factorization $\tilde T(z) = \prod_n P_n(z) B$ that converges uniformly on compact sets, where $B$ is a constant matrix.
\end{theorem}

\emph{The proof is given in Appendix~\ref{appendix:convergence}.}

\begin{remark}
Notice that $B$ in Theorem~\ref{theorem:convergence_general} is $J$-unitary and doubled-up. This follows  by noticing the product $\prod_n P_n(z)$ is $J$-unitary and doubled-up, and inverting it to get $B = \tilde T(z) \left(\prod_n P_n(z)\right)^{-1}$ and using Claims~\ref{claim:J_properties} and~\ref{claim:inverse_doubled_up}. 
\end{remark}

\subsection{Limiting behavior of $\tilde T(z)$}
\label{sec:static_squeezing}

For finite-dimensional systems, one way to characterize systems with no static squeezing was to examine the matrix $D$ in the state-space realization, and ensure it had the form $D = \Delta (\tilde D,0)$, where $\tilde D$ is a Hermitian matrix. Since the transfer function had the form $D + C(zI-A)^{-1}B$, we could take $|z| \to \infty$, and the direction along which the limit was taken did not matter.  However, as seen in \eqref{equ:limit_S_tilde}, when delayed feedback is present the direction of the limit is important, and can yield different results.  

Notice that obtaining different limit values of $\tilde T(z)$ as $|z| \to \infty$ is still consistent with the factorization $\tilde T(z) = \prod_n P_n(z) B$ of Theorem~\ref{theorem:convergence_general}, even though each term $P_n(z)$ approaches $I$ as $|z|\to\infty$ (along any direction). This is because the convergence of the product is only uniform on compact sets, and not the whole plane. Thus the limit and the product cannot be exchanged.

\section{Example}

\label{sec:example}

We study a modified case of an example network introduced in~\cite{gough2009enhancement}  and experimentally implemented in \cite{iida2012experimental}. In this network, a squeezer is placed in a feedback loop, resulting in enhanced squeezing. Our modification is the incorporation of a positive time delay in the feedback loop.
Another related example was constructed in \cite{crisafulli2013squeezed}, where a network of OPOs was interpreted  in terms of a quantum plant and a quantum controller. In that work, coherent feedback was shown to be capable of shifting the frequency of maximum squeezing and broadening the spectrum over a wider frequency band.
The effects of a time delay in coherent feedback networks on optical squeezing
 have  been studied in \cite{kraft2016time} and \cite{nemet2016enhanced}. In these works, it was demonstrated that the enhanced squeezing of quantum feedback networks may be improved by incorporating a time delay in the feedback loop. Depending on the implementation, the time delay can increase squeezing either on or off resonance. 

\begin{figure}[h]
\includegraphics[width=0.5\textwidth]{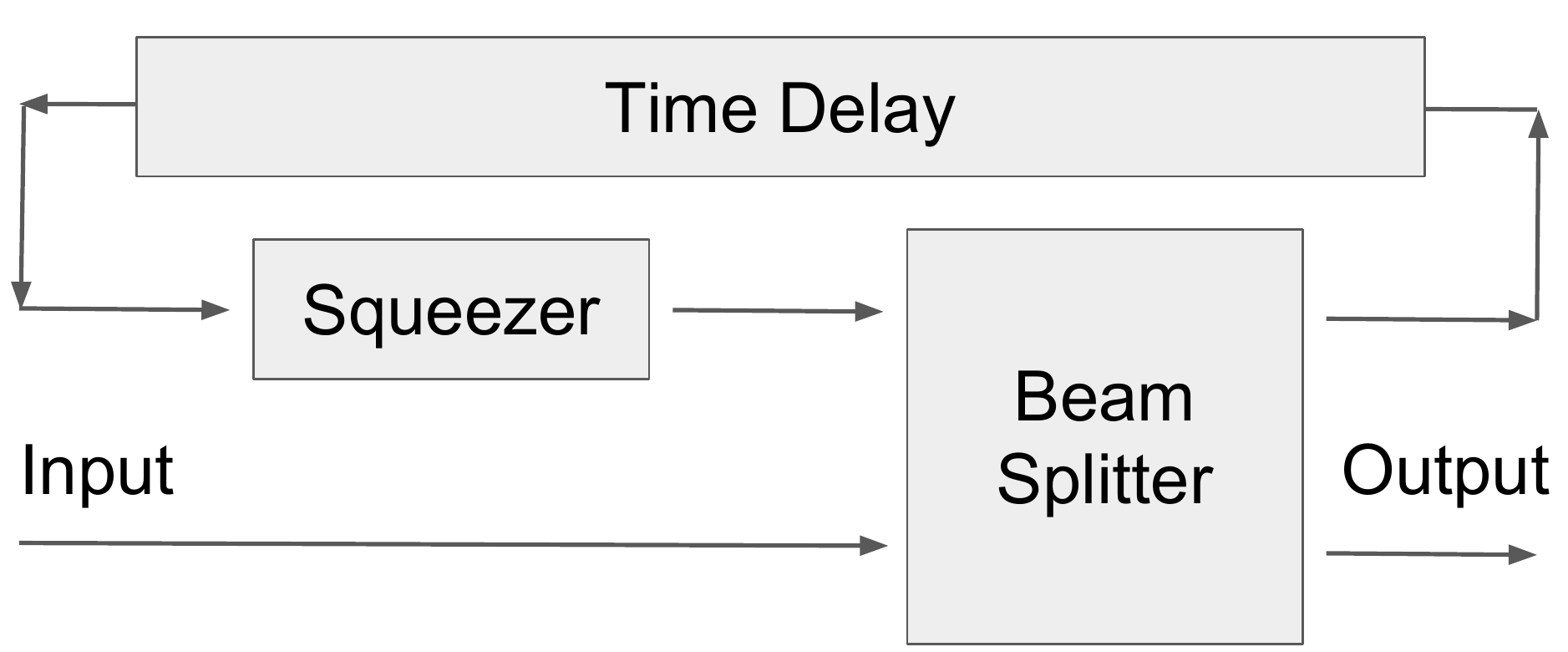}
\caption{{An example of a time-delay network in which enhanced squeezing occurs.}}
\label{fig:eg}
\end{figure}

\begin{figure}[h]
\includegraphics[width=0.5\textwidth]{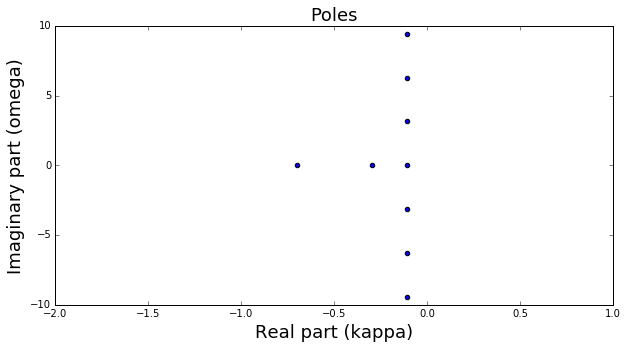}
\caption{{The poles of the transfer function of the example system}}
\label{fig:poles}
\end{figure}

For our construction, we start with the setup of Example VI.1 of~\cite{combes2017slh}.
Specifically, we begin with Eq.~(122) {from \cite{combes2017slh}}, in which the beamsplitter is placed in `series' behind the squeezer after an application of the feedback rule (with no delay). The SLH model is 
\begin{align}
G = (S,L,H) = \left(
\begin{pmatrix}
-\sqrt{1 - \eta^2} & \eta  \\
\eta & \sqrt{1 - \eta^2}
\end{pmatrix},
\begin{pmatrix}
-\sqrt{1 - \eta^2} \sqrt{\kappa} a \\
\eta  \sqrt{\kappa} a 
\end{pmatrix},
i \epsilon ({a^\dag }^2 - a^2)
 \right).
\end{align}
We find the transfer function from this SLH model, and add the time delay term $E(z) = \text{diag}(e^{-Tz}, e^{-Tz})$ as discussed in Section~\ref{sec:delayed_feedback}. The feedback loop is inserted from the first output to the first input.  The complete system is shown in Figure~\ref{fig:eg}. We set the parameters to be $T=2.0, \eta = 0.6, \kappa = 1.0, \epsilon = 0.2$.
%\red{[$\eta = 2$ is invalid, should have $\eta < 1$ here because it's the beamsplitter transmissivity.  And what is the value of $T$?]}.
The poles of the system are shown in Figure~\ref{fig:poles}.
There are two non-degenerate poles  (at $-\tfrac{1}{2}\kappa \pm \epsilon = -0.3, -0.7$) 
%\red{[One of your poles has positive real part, which means the system is unstable.  You're getting more net gain than loss per round trip.  Try to reduce the squeezing or increase the round-trip loss in the example to put all poles in the left half-plane.]} 
resulting from the squeezer, and one degenerate real pole (at about $-0.1$) along with an infinite set of complex poles resulting from the time delay. 

\begin{figure}[h]
\includegraphics[width=0.7\textwidth]{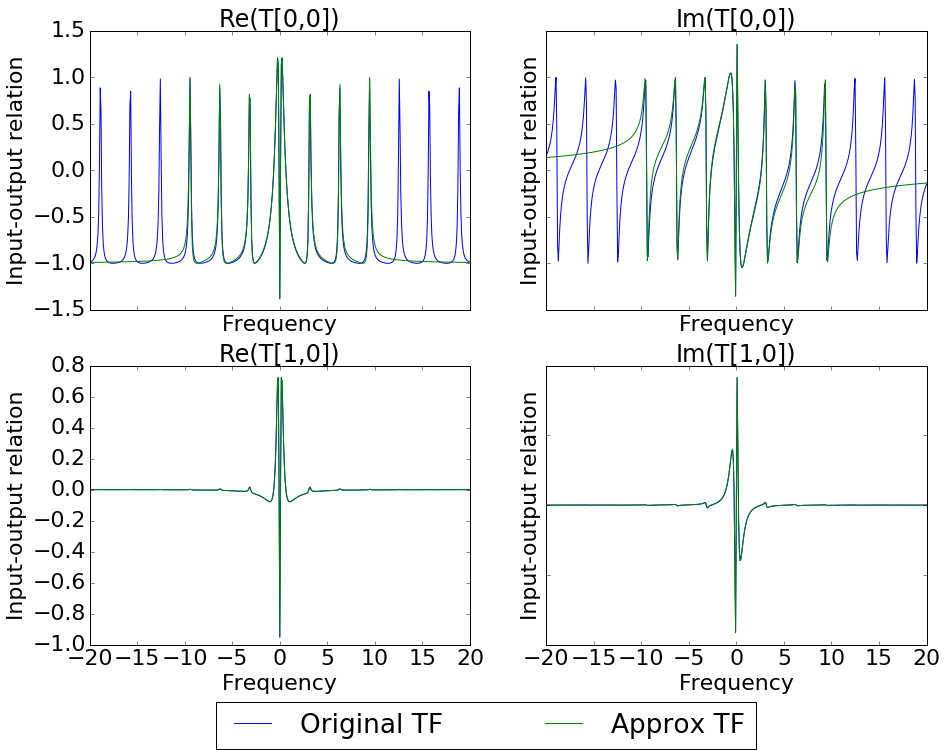}
\caption{%\red{[Make axis labels larger]}
 {The transfer function of the example system, and the transfer function of an approximation generated using the canonical factors. A constant pre-factor was added so that the two functions match at the origin. The real and imaginary parts of two components are shown along $z = i\omega$ for real values of $\omega$.}}
\label{fig:TF}
\end{figure}

The complex poles were degenerate in the sense that their eigenvectors did not satisfy the condition $v^\dag J v \ne 0$. 
Two ways to deal with this special case are (1) try to break the degeneracy by adding a phase shift within the loop and (2) perturb the eigenvector and use a modified factor.
When using (1), each complex pole and the degenerate real pole split into two complex poles. To obtain the factorization then, one would use the usual construction of complex poles, using every other pole as ordered in the imaginary direction (since each term also includes the conjugate pole, we do not include both nearly overlapping poles). When using (2), we use a modified factor, built of of two separate terms:
\begin{align}
P(z)= 
\left(I- VV^\flat + V \begin{pmatrix}
\frac{z - z_0}{z + \overline{z_0}} & 0 \\
0 & \frac{z - \overline{z_0}}{z + z_0}
\end{pmatrix}
V^\flat
\right)
\left(I- VV^\flat + V \begin{pmatrix}
 \frac{z - \overline{z_0}}{z + z_0} & 0\\
0 & \frac{z - z_0}{z + \overline{z_0}} 
\end{pmatrix}
V^\flat \right).
\end{align}
Each term corresponds to one of the terms resulting from the poles being perturbed.
To deal with the degenerate real pole using approach (2), one can find the two-dimensional eigenspace at the pole, and use both vectors to construct $V$ in the usual way for real poles.
%\red{[This confuses me.  The complex-pole eigenvectors are supposed to satisfy $v^\dagger J v = 0$.  There is no $v_1$ and $v_2$ (that condition is for real poles, Theorems 7.1-7.2).]}. 
For our example, we tried both approach (1) with a phase shift $\delta =  10^{-3}$ as well as approach (2) where the first entry of $v_1$ was increased by $\epsilon = 10^{-3}$. The resulting functions were visually indistinguishable when plotted along $z = i \omega$ for real values of $\omega$.
The original transfer function is shown against the approximated transfer function with the real poles and a total of six complex poles (or eight if the degenerate real pole is perturbed into two complex poles) in Figure~\ref{fig:TF}.
The code for the example is available on Github at \cite{Tabak2018}.

\section{Conclusion}

%\red{[Need to mention: this result holds for active systems in general.  2 x 2 blocks can include squeezing.]}

We have shown how a factorization theorem can be obtained linear quantum network with delayed feedback under some stated assumptions. 
The conditions are fairly general, and allow for a wide class of quantum linear systems which may be active.
The factorization generates a cascade of possibly infinitely many $2 \times 2$ quantum linear systems, with transfer functions $P_i(z)$. 
These canonical terms can also include squeezing.
Our theorem states that the factorization has form $\tilde T(z) = B \prod_i P_i(z)$, where $B$ is a constant matrix. We show that this product converges on compact sets. The $B$ matrix must be $J$-unitary and doubled-up as well.

One of our assumptions excludes cases with essentially feedforward delays, and other assumptions are made for simplicity. To get each individual factor, we also require non-degeneracy conditions to be met. However, as we showed in the example in Section~\ref{sec:example}, it is possible to find good approximations even when the non-degeneracy conditions are not met by perturbing the system.
%\red{[After redoing example with valid parameters, are the non-degeneracy conditions still broken?]}. 

\section{Acknowledgements}
GT is supported by an NDSEG fellowship awarded by the DoD and a Stanford University Graduate Fellowship. RH is supported by an appointment to the IC Postdoctoral Research Fellowship Program at MIT, administered by ORISE through U.S. DOE / ODNI. This work was also supported by ARO under grant ``W911NF-16-1-0086.''

We would also like to thank discussions with Tatsuhiro Onodera, Nikolas Tezak, Hendra Nurdin, John E. Gough, Edwin Ng, Daniel B. Soh,  Niels Lorch, and Eric Chatterjee.

\appendix
\section{Proof of Lemma~\ref{lemma:static_approximation}}
\label{appendix:approx}

\begin{remark}
Because of the relationship we found between the zeros and poles for $J$-unitary functions (along with the eigenvectors at the zeros and poles), the lemma holds for the roots if and only if it holds for the poles. Therefore, it suffices to prove the lemma only for one or the other.
 {We will prove parts (i) and (iii) of the theorem for the poles and part (ii) for the zeros.}
 \end{remark}

\emph{proof for (i):} 

We begin by showing that for sufficiently large $M$, for each zero $w_0$ of $f_S(z)$ such that $|w_0| > M$, there is a radius $\epsilon_M = O(1/M)$ such that exactly one zero $z_0$ of $f_T(z)$ is within $\epsilon_M$ of $w_0$. To show this, we will use  Rouch\'{e}'s Theorem~\cite{saff1976fundamentals}. This theorem states that if $| f_T(z) - f_S(z) | < |f_T(z)|, |f_S(z)|$ for every $z$ on the boundary of some Jordan curve $\Gamma$, then $f_T(z)$ and $f_S(z)$ will have the same number of zeros in the interior of $\Gamma$. 

We can ensure that for sufficiently large $M$, each $\epsilon_M$ is small enough so that it isolates a single zero $w_0$ of $f_S(z)$. 
{
To do this, first note that there is always a positive upper bound $\epsilon_\text{max}>0$ for the radii that isolate the zeros  and poles of a  not identically zero meromorphic function from  other zeros and poles on a bounded set.
Using the periodicity of $f_S(z)$ and the boundedness of the real part of the the zeros (Lemma~\ref{lemma:poles_f}), we can get the desired result.} 
Given a relation $\epsilon_M = O(1/M)$ for sufficiently large $M$, we can make $\epsilon_M$ as small as we like (in particular we can satisfy $\epsilon_M < \epsilon_\text{max}$) by requiring $M$ to be sufficiently large.

In order to pick the bounds for Rouch\'{e}'s theorem,  we expand $f_S(z)$ at $w_0$. Explicitly, assuming that leading order terms have order one (see Remark~\ref{remark:rate_of_convergence}), we can expand:
\begin{align}
\label{equ:fS_expansion}
f_S(z) = c_1^{(S,w_0)}(z-w_0) + c_2^{(S,w_0)}(z - w_0)^2 + ...
\end{align}
%We ensure $\epsilon$ is small for each $w_0$ so that if  $z \in B_\epsilon(w_0)$ then $|z| >|w_0| - \epsilon = O(|w_0|)$ (e.g. $\epsilon < w_0/2$). This bound ensures that there is a $\delta_M$ for which $|f_S(z) - f_T(z)| \le \delta_M = O (1/M)$ for $z \in B_\epsilon(w_0)$.
%  $f_S(z) = c_1^{(w_0)} (z - w_0) + O(\epsilon^2),$ so t
There is an error bound $k^{(S,w_0)} > 0$ such that if $z \in B_\epsilon(w_0)$, then
  \begin{align}
  \label{equ:k_s}
 |f_S(z) - c_1^{(S,w_0)} (z - w_0) | < k^{(S,w_0)} \epsilon^2.
 \end{align}
%   Suppose that $|f_S(z) - f_T(z)| \le \delta_M = O(1/M)$ on the boundary of the $\epsilon$-ball around $w_0$. 
  From this we obtain the bounds for $z$ on $\partial B_\epsilon(w_0)$,
  \begin{align}
  |f_S(z)| &>  |c_1^{(S,w_0)}| \epsilon - k^{(S, w_0)} \epsilon^2, \\
    |f_T(z)| &>  |c_1^{(S,w_0)} | \epsilon - \delta_M -k^{(S,w_0)} \epsilon^2 . \label{equ:fT_bound}
  \end{align}
  Above in Eq.~(\ref{equ:fT_bound}), we introduce $ \delta_M = \sup_{\partial B_\epsilon(w_0) } |f_S(z) - f_T(z)| $. Notice sufficient conditions for Rouch\'{e}'s theorem are now $\delta_M < |f_S(z)|, |f_T(z)|$.
  
From Lemma~\ref{lemma:strip}, the poles of $\tilde S(z)$ are contained in a strip of bounded $\Re (z)$. This implies that there is a bound on the real part of all zeros of $f_S(z)$.
  Using Lemma~\ref{lemma:rate_of_convergence} (ii), we can bound $\delta_M = O(1/M)$  for large enough $M$.
  We can also bound the coefficients $c_1^{(S,w_0)}$ for all zeros $w_0$ of $f_S(z)$ from below by $C > 0$. This follows from the periodicity of $f_S(z)$ in the imaginary direction, and noting that the bound on the real part of $z$ for the zeros of $f_S(z)$ implies that in each period there are a finite number of zeros.
   Similarly, we can bound all $k^{(S,w_0)}$ from above by a constant $k$. 
   If we pick $\epsilon_M = \frac{3 \delta_M}{C}  = O(1/M)$ for sufficiently large $M$, the desired conditions for Rouch\'{e}'s theorem follow from the above inequalities. 

For the case where $f_S(z)$ and $f_T(z)$ are exchanged, a small modification is needed to obtain the desired $\epsilon_M$, since  $f_T(z)$ may not be periodic. 
{
First, we need to modify $\epsilon_{\text{max}}$ to ensure the poles $z_0$ are isolated. This can be done by an application of the triangle inequality.
}
Next, we are interested in bounds for the constants $c_1^{(T,z_0)}$ and $k^{(T,z_0)}$  when expanding $f_T(z)$ at each of its zeros $z_0$, analogous to the constants used above in Eq.~(\ref{equ:fS_expansion}-\ref{equ:k_s}). Using Cauchy's integral formula on the $\epsilon$-balls in the above arument, and the error estimate $|f_T(z) - f_S(z)| < \delta_M$, we can obtain the same bounds $C$ and $k$ above for the expansion coefficients $c_i^{(T,w_0)}$ and $k^{(T,z_0)}$, respectively, up to an error of $\delta_M = O(1/M)$. Therefore the desired result holds in this case as well.
$\blacksquare$

% Suppose $c_^{(T,z_0)}$ and $c_S^{(i)}$ are the coefficients of the  series expansions of $f_T(z)$ and $f_S(z)$, respectively, at points such that $|z| > 1/\epsilon$. Then using Cauchy's differentiation formula, one can arrive at $|c_T^{(i)} - c_S^{(i)}| = O(\epsilon)$. From this, a controllable error term can be added in the analysis above, leading to the desired result.

% $| f_T(z) - f_S(z) | < |f_T(z)|, |f_S(z)|$ for $z$ on the boundary of a neighborhood outside $|z| > M$ and around a zero $z_0$ of $f_T(z)$ or $w_0$ of $f_S(z)$ (which we can make as small as we like, depending on $M$). We can guarantee the neighborhood has only one zero for the given function by making it small enough.
 
% (ii)
%The orthogonal projectors can be made to satisfy $\| P_T - P_S \| < \epsilon$ for sufficiently large $|z|$ since projectors to the eigenspace of a matrix are continuous within some neighborhood~\cite{kato1976grundlehren}. Such a bound can be made to hold uniformly for the zeros (poles) with large enough $|z|$, since $\tilde S(z)$ is periodic.
\emph{proof for (ii):} 

Suppose we are given a pair of roots from part (i), $z_0$ for $f_T(z)$ and $w_0$ for $f_S(z)$, along with a radius $\epsilon_M$
such that $z_0$ is the only root of $f_T(z)$ and $w_0$ is the only root of $f_S(z)$ inside $B_{\epsilon_M}(w_0)$. 
 From $z_0$ and $w_0$ (which are poles of $\tilde T(z)$ and $\tilde S(z)$) we can find the corresponding roots of $\tilde T(z)$ and $\tilde S(z)$ (from Claim~\ref{claim:vectors_zeros_poles}), which we label $\hat z_0$ and $\hat w_0$, respectively. Below we will consider all such possible pairs $\hat z_0$ and $\hat w_0$.

We will show first that 
\begin{align}
\| \tilde T(\hat z_0) - \tilde S(\hat w_0) \|  = O (\epsilon_M).
\label{equ:pole_bound}
\end{align}
To do this, write
\begin{align}
\| \tilde T(\hat z_0) - \tilde S(\hat w_0) \| 
\le \| \tilde T(\hat z_0) - \tilde S(\hat z_0) \|  + \| \tilde S(\hat z_0) - \tilde S(\hat w_0) \|.
\end{align}
To find a bound for $\| \tilde T(\hat z_0) - \tilde S(\hat z_0) \|$ independent of the choice of $\hat z_0$ for sufficiently large $|z|$, we can use the bound $\kappa_M = \sup_{|z| > M} \|T_i(z) - S_i \| = O(1/M)$ for $i = 1,2,3,4.$

We can bound
\begin{align}
& \| \tilde T(z) - \tilde S(z) \| 
\le \|T_1(z) - S_1\| \\
&+ \| T_2(z) - S_2 \|
\| E(z)\|
\| (I - T_4(z) E(z))^{-1} - (I - S_4 E(z))^{-1}  \|
\| T_3(z) - S_3 \| \nonumber
\end{align}
The $E(z)$ term is bounded for all $\hat{z}_0$, following Lemma~\ref{lemma:rate_of_convergence} (ii).
The term for which obtaining the bound is not obvious is
$\|(I - T_4(z) E(z))^{-1} - (I - S_4 E(z))^{-1}\|$. We do this below.
When $|z|$ is sufficiently large, we can use Lemma~\ref{lemma:rate_of_convergence} to write $T_4(z) = S_4 + \nu_{\hat w_0} D_{\hat w_0}(z)$  on each $B_{\epsilon_M}(\hat w_0)$ where all the functions $D_{\hat w_0}$ are bounded and matrix-valued, and $\nu_{\hat w_0} = O(1/|{\hat w_0}|)$.
When $z \in B_{\epsilon_M}(\hat w_0)$,
\begin{align}
&(I - T_4(z) E(z))^{-1} \label{eq:approx_T_S} \\
 & = (I - S_4 E(z) - \nu_{\hat w_0} D_{\hat w_0} E(z) )^{-1} \nonumber \\
 & = (I - \nu_{\hat w_0} (I - S_4 E(z ))^{-1} D_{\hat w_0} E(z) )^{-1}  (I - S_4 E(z))^{-1} \nonumber \\
 & = ( I + O(\nu_{\hat w_0}) )^{-1} (I - S_4 E(z))^{-1} \nonumber.
\end{align}
{In the final step of \eqref{eq:approx_T_S},} we notice that since $\epsilon_M$ isolates the  zero $\hat w_0$ from other zeros and poles of $\tilde S(z)$, the function $ (I - S_4 E(z))^{-1} $ is bounded on $B_{\epsilon_M}(\hat w_0)$.  From this we find
\begin{align}
&(I - T_4(z) E(z))^{-1} - (I - S_4 E(z))^{-1} \\
&= [( I + O(\nu_{\hat w_0}) )^{-1}-I] (I - S_4 E(z))^{-1} \nonumber \\
&=  ( I + O(\nu_{\hat w_0}) )^{-1} O(\nu_{\hat w_0}) (I - S_4 E(z))^{-1} \nonumber
=O(\nu_{\hat w_0}).
\end{align}
Since $\hat z_0$ is in $B_{\epsilon_M}(\hat w_0)$, This gives the desired bound
\begin{align}
\label{equ:bound_1}
\| \tilde T(\hat z_0) - \tilde S(\hat z_0) \| = O(1/M).
\end{align}
 Further, by a periodicity argument similar to that in part (i), we can make the bound independent of the choice of zeros $\hat z_0$ and $\hat w_0$.

Next, we obtain a similar bound for  $\| \tilde S(\hat z_0) - \tilde S(\hat w_0) \|$:
\begin{align}
 \| \tilde S(\hat z_0) - \tilde S(\hat w_0) \|
 \le \| S_2 \| \|E(z) \| \|(I - S_4 E(\hat z_0))^{-1} - (I - S_4 E(\hat w_0))^{-1}\| \|	S_3\|.
\end{align}
We notice that $\|E(\hat z_0) - E(\hat w_0) \| = O (\epsilon_M)$, so using a similar argument as before,
\begin{align}
(I - S_4 E(\hat z_0))^{-1}
= (I - S_4 E(\hat w_0))^{-1} (I + O (\epsilon_M)).
\end{align}
We thus obtain 
\begin{align}
\label{equ:bound_2}
\| \tilde S(\hat z_0) - \tilde S(\hat w_0) \|= O(1/M)
\end{align}
Using Eq.~(\ref{equ:bound_1}) and Eq.~(\ref{equ:bound_2}), we finally obtain  Eq.~(\ref{equ:pole_bound}).

Next, to relate our analysis to the projectors, we can use the formula~\cite{kato1976grundlehren}
\begin{align}
P = -\frac{1}{2 \pi i} \int_\Gamma R(\zeta, A)  d \zeta,
\end{align}
where $ R(\zeta, A) = (A- \zeta)^{-1}$ is the resolvent of a matrix $A$, and $P$ is the sum of the eigenprojections of all the eigenvalues of $A$ inside the contour $\Gamma$. 

We will use the second resolvent identity below. It states that for two matrices $A$ and $B$, and $\zeta$ a number not in the spectrum of either, 
\begin{align}
R(\zeta,A) - R(\zeta,B) = R(\zeta,A)(B-A) R(\zeta,B).
\end{align}

In particular, since $\hat z_0$ and $\hat w_0$ are roots of $\tilde T(z)$ and $\tilde S(z)$, respectively, the matrices $\tilde T(\hat z_0)$ and $\tilde S(\hat w_0)$ both have the eigenvalue zero. For a sufficiently small contour $\Gamma$, the zero eigenvalue becomes isolated for both $\tilde T(\hat z_0)$ and $\tilde S(\hat w_0)$.
{The curve $\Gamma$ can be made a small, but fixed size for all $\hat w_0$ and $\hat z_0$. This will allow us to bound the resolvent $R$ below to be bound by a constant.
}
 From the second resolvent identity, we can obtain the bound:
\begin{align}
\|P_T - P_S \| \le \frac{|\Gamma|}{2 \pi } \sup_{\zeta \in \Gamma} R(\tilde T(\hat z_0),\zeta)  \sup_{\zeta \in \Gamma} R(\tilde S(\hat w_0),\zeta) \| \tilde T(\hat z_0) -  \tilde S(\hat w_0)\| = O(1/M).
\end{align}
Above, $|\Gamma|$ indicates the length of the curve $\Gamma$. 
%\red{Missing some terms in the above equation, namely $2\pi$ and the length of $\Gamma$.  Also, need to emphasize that the curves $\Gamma$ have fixed size (set by $\epsilon_{\rm max}$ I think), so $R$ is bounded by a constant.}
The bound  is independent of the choice of $\hat w_0$ and $\hat z_0$.
This completes the proof.

$\blacksquare$

\emph{proof (iii)}
The proof follows the first step used to prove (ii),  {using the poles instead of the zeros. The boundedness of $(I-S_4E(z))^{-1}$ is obtained by the fixed bound $\epsilon$ away from the poles.}
$\blacksquare$

\begin{remark}
Note on the rate of convergence:
\label{remark:rate_of_convergence}
If  the zeros have order $p$ instead of $1$, the scaling of $\epsilon_M$ follows $O(M^{-1/p})$.
This can be seen from the order of the terms required in the series expansion used to prove part (i).
\end{remark}

\section{Proof of Theorem~\ref{theorem:convergence_static}}
\label{appendix:convergence_static}

The roots come in conjugate pairs (Claim~\ref{claim:doubled_zeros}), so the set of roots can be partitioned into two conjugate sets: $Z = Z_+ \cup Z_-$ where $Z_- = \{ \overline z : z \in Z_+\}$.  Since we assumed the delays are commensurate, the function $\tilde S(z)$ is periodic along the imaginary axis with some period $P$.  Thus the roots in $Z_+$ (and $Z_-$) can be ordered as:
\begin{equation}
	Z_+ = \Bigl\{ \underbrace{z_m + i P n}_{z_{m,n}} :  m \in \{1,...,M \}, n \in \mathbb{Z} \Bigr\}
\end{equation}
where $M$ is the number of roots (in $Z_+$) per period; because roots come in conjugate pairs, in total there are $2M$ roots per period.  To satisfy the condition that all roots are simple (Assumption (\ref{assumption:simple}) of Section~\ref{sec:assumptions}), we require that the $z_n$ be unique, satisfying $0 < {\rm Im}(z_m) < P$ and ${\rm Im}(z_m) \neq P/2$.  Thus all roots are complex, and a decomposition of the form (\ref{equ:OPO_complex}) will be possible.  We choose the ordering:
\begin{align}
\label{equ:S_expand}
\tilde S(z) = \Bigl(\prod_{m=1}^K \Bigl[\prod_{n=-\infty}^\infty P_{m,n}(z)\Bigr]\Bigr) B(z) .
\end{align}
Here, $P_{m,n}(z)$ is the elementary factor with zeros $(z_{m,n}, \overline{z}_{m,n})$ and poles  $(-z_{m,n}, -\overline{z}_{m,n})$ defined in Eq.~(\ref{equ:OPO_complex}).

We first show that each infinite product in Eq.~(\ref{equ:S_expand}) converges uniformly on bounded sets.  Starting with $m=1$, consider the terms $P_{1,1}, P_{1,2}, \ldots$ generated by detaching the zeros $z_{1,1}, z_{1,2}, \ldots$ individually from $\tilde{S}(z)$.  Since $\tilde S(z)$ is periodic with period $P$, so the projector of $\tilde S(z)$ at $z_{m,n}$ is independent of $n$.  Thus the projectors in the terms used to detach the terms $P_{1,1}, P_{1,2}, \ldots$ are the same, and these terms commute.  Therefore, these terms can be used to sequentially detach the roots of $\tilde S(z)$, giving the expression:
\begin{align}
	\prod_n P_{m,n}(z) 
%	& = \prod_n \left[I - V_m V_m^\flat + V_{m} 
%		\begin{pmatrix} \frac{z - z_{m,n}}{z + \overline{z}_{m,n}} && 0 \ 0 && {a_{m,n}(\overline z)}^* \end{pmatrix}
%		V_m^\flat \right] \\
	& = I - V_m V_m^\flat 
		+ V_m \prod_n \begin{pmatrix} \frac{z - z_{m,n}}{z + \overline{z}_{m,n}} & 0 \\ 0 & \left(\frac{\overline{z} - z_{m,n}}{\overline{z} + \overline{z}_{m,n}}\right)^* \end{pmatrix} V_m^\flat.
\end{align}
where we have fixed $m = 1$ for the moment.  

The above product converges because one can show \cite[Appendix E]{tabak2016trapped} that the following series converges uniformly on bounded sets:
\begin{align}
	Q_m(z) & \equiv \prod_{n=-\infty}^{\infty} \frac{z - z_{m,n}}{z + \overline{z}_{m,n}}
	= \frac{\sinh\bigl(\frac{\pi}{P}(z - z_m)\bigr)}{\sinh\bigl(\frac{\pi}{P}(z + \overline{z}_m)\bigr)} \label{eq:qsinh}
\end{align}
After the zeros $z_{1,1}, z_{1,2}, \ldots$ are detached, we obtain a function {$\bigl(\prod_i P_{1,i}(z)\bigr)^{-1} \tilde S(z)$}
% \red{(I think this is correct.  Please check.)}
   which is again periodic in the imaginary direction with period $P$, but has only $2(M-1)$ zeros per period.  We can repeat the same procedure for the remaining sequences of zeros $z_{m,1},z_{m,2},...$ in $Z_+$, finding:
\begin{equation}
	\tilde{S}(z) = \prod_{m=1}^K \biggl[1 - V_m V_m^\flat 
		+ V_m \prod_n \begin{pmatrix} Q_m(z) & 0 \\ 0 & Q_m(\overline{z})^* \end{pmatrix} V_m^\flat \biggr] B(z).
\end{equation}
$B(z)$ is an entire function that, like $\tilde{S}(z)$ and $Q_m(z)$, is periodic along the imaginary axis with period $P$.  In the limits ${\rm Re}(z) \rightarrow \pm\infty$, we can show that $B(z)$ tends to a constant by examining the behavior of $\tilde{S}(z)$ and $Q_m(z)$.  For $\tilde{S}(z)$, recall that it is given by (\ref{equ:S_tilde}):
\begin{align}
	\tilde{S}(z) & = S_1 + S_2  E(z) \left( I - S_4 E(z) \right)^{-1} S_3 \nonumber \\
		& = S_1 + S_2  \left( E(-z) - S_4 \right)^{-1} S_3
\end{align}
Using the fact that $S_4$ is invertible and $E(z) \rightarrow 0$ for ${\rm Re}(z) \rightarrow +\infty$, it is easy to find the limiting behavior of $\tilde{S}(z)$.  The limits of $Q_m(z)$ for ${\rm Re}(z) \rightarrow \pm\infty$ are found using (\ref{eq:qsinh}) and the asymptotic relation $\sinh(x + iy) \rightarrow \tfrac{1}{2}{\rm sgn}(x) e^{|x| + iy\, {\rm sgn}(x)}$:
\begin{align}
\label{equ:limit_S_tilde}
	\tilde{S}(z) & \rightarrow \begin{dcases}
	S_1 & {\rm Re}(z) \rightarrow +\infty \\
	S_1 - S_2 S_4^{-1} S_3  & {\rm Re}(z) \rightarrow -\infty
	\end{dcases}, \\
	Q_m(z) & \rightarrow \begin{dcases}
	e^{-(2\pi/P){\rm Re}(z_m)} & {\rm Re}(z) \rightarrow +\infty \\
	e^{+(2\pi/P){\rm Re}(z_m)} & {\rm Re}(z) \rightarrow -\infty
	\end{dcases}.
\end{align}
These limits prove that $B(z)$ tends to a constant for $z \rightarrow \pm\infty$, when $\Im(z)$.
 {We also know that $B(z)$ is periodic in the imaginary direction, since it is the product of commensurate periodic functions. Combining these two results, we determine that $B(z)$ has bounded range on the entire plane.
  By Liouville's theorem, it must be a constant.  }
%\red{Need to invoke periodicity in imaginary direction for this to work, because we only proved boundedness in the limit ${\rm Re}(z) \rightarrow \pm \infty$.}

$\blacksquare$

\section{Proof of Theorem~\ref{theorem:convergence_general}}
\label{appendix:convergence}

The intuition for the proof of convergence  will be to approximate the function $\tilde T(z)$ by $\tilde S(z)$ for large $|z|$ using Lemma~\ref{lemma:static_approximation}. 

From Lemma~\ref{lemma:static_approximation}, we intuitively have that the zeros (poles) of $\tilde T(z)$ and $\tilde S(z)$ become arbitrarily close to one another as $|z| \to \infty$.
Using this lemma, one can find a ball $B_{R_0}(0)$, with the following property:
 There is a one-to-one correspondence between the zeros (poles) of $\tilde T(z)$ and $\tilde S(z)$ outside  $B_{R_0}(0)$, such that a zero (pole) $z_0$ of $\tilde T(z)$ and a zero (pole) $w_0$ of $\tilde S(z)$ correspond to each other if and only if $|z_0 - w_0| < \epsilon_{R_0} = O(1/R_0)$. 
%\item $|R - |\Im w_0| |  > \epsilon_{R_0}$ for all roots and poles of $\tilde S(z)$. 
%Property 2  ensures that all squares $\mathcal{S}_i$ for $i = 1,2,...$ with the center at the origin and half-length $R_i = R_0 + i P$ have properties 1 and 2 (with the appropriate subscript $i$ replacing $0$), where $P$ is the periodicity of the roots of $\tilde S(z)$ in the imaginary direction.

The order of terms being detached from $\tilde T(z)$ will be determined as follows.
 For terms with zeros inside $B_{R_0}(0)$, the order is not important for convergence, 
 since there are a finite number of zeros inside $B_{R_0}(0)$.
 For example, we can pick the zeros according to increasing norm. 
For terms with zeros outside $B_{R_0}(0)$, the order is determined using the order of zeros of $\tilde S(z)$ in Theorem~\ref{theorem:convergence_static}.  That is, we use the one-to-one correspondence between the zeros of $\tilde S(z)$ and $\tilde T(z)$ to determine the order of zeros of $\tilde T(z)$. Detaching terms in this order gives $K$ (possibly) infinite products  {$\prod_n P_{m,n}$ for $m = 1,...,K$}, where $K$ is the number of zeros in each period of the zeros of $\tilde S(z)$ in Theorem~\ref{theorem:convergence_static}. Our task is to show the products  {$\prod_n P_{m,n}$}  converge {for each $m$}. 

Consider one of the products  {$\prod_n P_n (z)$} (dropping  {the index m}). Suppose the corresponding product generated for $\tilde S(z)$ by Theorem~\ref{theorem:convergence_static} is  {$\prod_n P_n^{S}(z)$}. Here we do not include the terms with zeros inside $B_{R_0}(0)$.
It will be convenient to use the form of the elementary factors introduced in Section~\ref{sec:another_form}.
We can write
\begin{align}
\label{equ:Fn_terms}
P_n^{S}(z) = M F_n^S(z) M^\dag J, &&
F_n^S(z) = \text{diag} (a_{w_n}(z), -a_{w_n}(z^*)^*, 1,-1,...,1,-1).
\end{align}
 where $a_{z_0}(z) = \frac{z - z_0}{z + z_0^*}$ and $M$ is a fixed $J$-unitary. Here  {$w_n$} are the appropriate zeros of $\tilde S(z)$.  {Above in Eq.~(\ref{equ:Fn_terms}), M is fixed due to the periodicity argument used in Section~\ref{theorem:convergence_static}}.
The terms  {$P_n(z)$} (with the zeros of $\tilde T(z)$) have form 
{
\begin{align}
P_n(z) &= \tilde M_n F_n(z) \tilde M_n^\dag J, \\
F_n(z) &= \text{diag} (a_{z_n}(z), -a_{z_n}(z^*)^*, 1,-1,...,1,-1).
\end{align}}
Here {${z_n}$} are the appropriate zeros of $\tilde T(z)$, and  { $ \tilde M = {M + \epsilon_{R_n} N_n} $} is $J$-unitary.
The {$N_n$} are bounded matrices and {$\epsilon_{R_n} = O(1/R_n)$}. 
 {In order to produce the $\tilde M_n$ related to $M$ by error $\epsilon_{R_n}$,  apply  Lemma~\ref{lemma:static_approximation}(ii) to relate the $W$ matrices of $P_n$ and $P_n^{S}$ in the form of \eqref{equ:W_form}. In the remainder of the procedure of Section~\ref{sec:another_form}, we complete the basis of each to produce the respective $M$ matrices of \eqref{equ:J_unitary_form}. Care must be taken to ensure this procedure maintains the error bound for the $M$ matrices. For example, we can use Gram-Schmidt (with the indefinite basis), appending the same set of initial vectors in both cases. }

Since we are interested in the convergence of  {$\prod_n P_n (z)$} on compact sets, fix $r>0$.
We will show  {$\prod_n P_n(z)$}   converges uniformly on $B_r(0)$.
We will approximate  {$\prod_n P_n(z)$} on $B_r(0)$ using a sequence of functions  {$T_\ell(z) = \bigl( \prod_{n = \ell+1}^\infty P_n(z)\bigr)$} for  {$\ell = 0,1,...$}. 
{Notice that  $\prod_n P_n(z) = \left(\prod_{n=1}^{\ell} P_n(z)\right) T_\ell(z)$. Intuitively, we can think of the $T_\ell(z)$ as the truncation error. If the error can be controlled, then we can obtain a convergence result. Towards our goal we will show the following claim.}
\begin{claim}
\label{claim:cauchy_conv}
 { $T_\ell(z)$} converges uniformly on $B_r(0)$ {as $\ell\to\infty$}, using the supremum norm.  
 \end{claim}
%\red{(I think you need to say ``compact sets that exclude all poles $(-z_n^*)$''.  Because $T_\ell(z)$ and $T_\infty(z)$ differ by a product: $T_\ell(z) = T_\infty(z) \bigl(\prod_{n=\ell+1}^{\infty} P_n(z)\bigr)^{-1} \prod_{n=\ell+1}^{\infty} P_n^S(z) \equiv T_\infty(z) W_\ell(z)$ (I'm calling it $W$ because not used elsewhere).  You show below that $\lim_{\ell\rightarrow\infty} ||W_\ell(z) - 1|| = 0$ uniformly on bounded sets below, which is good.  But since $T_\infty(z)$ has poles inside $B_r(0)$, you won't be able to prove convergence in the vicinity of the poles, because $T_\ell(z) - T_\infty(z) = (W_\ell(z) - 1) T_\infty(z)$ also has poles.)}

To prove the above claim, we will use our knowledge of  {$P_n(z)$} and  {$P_n^S(z)$} to show that for sufficiently large  {$R_n$},  {$\|P_n(z) - P_n^S(z)\| = O(1/R_n^2)$} for $z \in B_r(z)$ where the zeros of  {$P_n(z)$} and  {$P_n^S(z)$} are outside  {$B_{R_n}(0)$}.
 {First, we show an estimate of the same order for}  {$F_n(z)$} and  {$F_n^S(z)$}:
 {
\begin{align}
\label{equ:Fs_esimate}
\|F_n(z) - F_n^S(z)\|
\le \|{F_n^S(z)} \| \| {F_n^S(z)}^{-1} F_n(z) - I \|
\end{align}
}
 {
For large enough $R_n$, we can use $|a_{w_n}(z) - 1| = O(1/R_n)$ for $|w_n| > R_n$ to show
\begin{align}
\label{equ:Fs_norm_estimate}
  {F_n^S(z)}  = J + O(1/R_n).
 \end{align}
 }
Here we use the rate of convergence result of  Lemma~\ref{lemma:static_approximation}.
For estimating the term $\| {F_n^S(z)}^{-1} F_n(z) - I \|$, one can compute (using $\gamma_n = w_n - z_n = O(1/R_n)$):
\begin{align}
a_{w_n}(z) a_{z_n}(z)^{-1} &= 
\frac{z - w_n}{z + w_n^*}
\frac{z+z_n^*}{z - z_n} =
\frac{z - (z_n + \gamma_n)}{z + (z_n+ \gamma_n)^*}  \frac{z + z_n^*}{z - z_n} =
\frac{1- \frac{\gamma_n}{z-z_n}}{1 + \frac{\gamma_n^*}{z + z_n^*}} \label{equ:estimate_a_terms} \\
& = \left(1 - \frac{\gamma_n}{z - z_n}\right)
\left(1 - \frac{\gamma_n^*}{z + z_n^*}\right)
+ O(|\gamma_n|^2) \nonumber \\
&= 1 - \frac{\gamma_n}{z - z_n}
- \frac{\gamma_n^*}{z + z_n^*}
+ O(|\gamma_n|^2) \nonumber \\
&= 1 - \frac{\gamma_n}{z - w_n}
- \frac{\gamma_n^*}{z + w_n^*}
+ O(|\gamma_n|^2). \nonumber
\end{align}
For large enough $R_n$, the terms $(z- z_n)^{-1} = O(1/R_n)$ and $ (z + z_n^*)^{-1} = O(1/R_n)$ since $|z_n| >R_n$.
 {
Thus we find
\begin{align}
\label{equ:a_a_inv_estimate}
a_{w_n}(z) a_{z_n}(z)^{-1} &= 1 + O(|1/R_n^2|).
\end{align}
}
 {Using \eqref{equ:Fs_norm_estimate} and \eqref{equ:a_a_inv_estimate} in \eqref{equ:Fs_esimate}, we find} 
\begin{align}
\|F_n(z) - F_n^S(z)\|
 = (1 + O(1/R_n)) O(1/R_n^2) = O(1/R_n^2).
 \end{align}
We can then compute
\begin{align}
\|P_n(z) - P_n^S(z)\| 
  &= \| (M+ \epsilon_{R_n} N_n) F_n(z) (M + \epsilon_{R_n} N_n)^\dag J
   - M F_n^S(z) M^\dag J \|  \nonumber \\
  &= \| (M+ \epsilon_{R_n} N_n) F^S_n(z) (M + \epsilon_{R_n} N_n)^\dag J
   - M F_n^S(z) M^\dag J \| + O(1/R^2) \nonumber \\
   &= \epsilon_{R_n} \| N_n  F^S_n(z) M ^\dag  J+ M F^S_n(z) N_n^\dag  J \| + O(1/R_n^2).
   \label{equ:estimate_order_eps}
 \end{align}
Using that $M$ and  $M+ \epsilon_{R_n} N_n$ are $J$-unitary, we find
\begin{align}
(M+ \epsilon_{R_n} N_n)J(M+ \epsilon_{R_n} N_n)^\dag - MJM^\dag = 0,
\end{align}
from which we find
\begin{align}
\label{equ:matrix_estimate_squared}
N_n  J M^\dag + M J N_n^\dag = O(1/R_n^2).
\end{align}
Using $F_n^S(z) = J + O(1/R_n)$, \eqref{equ:estimate_order_eps}, and \eqref{equ:matrix_estimate_squared}, we find
\begin{align}
\label{equ:quad_bound}
\|P_n(z) - P_n^S(z)\| = O(1/R_n^2).
\end{align}

This gives us a bound on the estimate error between $P_n(z)$ and $P_n^S(z)$ inside $B_r(0)$ for some sufficiently large $R_n$, which will be key in showing the convergence of $T_\ell(z)$. 
Notice that because the zeros $w_n$ of $\tilde S(z)$ are periodic, we can replace $O(1/R_n^2)$ by $O(1/n^2)$. Let $q$ be the smallest index {of $n$ for which the above estimates Eqs.~(\ref{equ:Fs_norm_estimate}, \ref{equ:quad_bound}) hold}, and also that $R_n > r$.

The next step will be to show that $\|\prod_{n=q}^p P_n(z)\|$ is bounded in $B_r(0)$ for all $p = q,q+1,...$.  {Our estimate \eqref{equ:quad_bound} produces}
\begin{align}
\label{equ:bounded_P}
\left\|\prod_{n=q}^p P_n(z)\right\| \le 
\left\| \prod_{n=q}^p P_n^S(z) \right\| \prod_{n=q}^p \left( 1 +  \frac{c}{n^2 \|P_n^S(z)\|} \right).
\end{align}
The product $ \prod_{n=q}^p P_n^S(z) $ is bounded on $B_r(0)$  {over all} $p = q,q+1,...$ since it converges in the limit $p \to \infty$, and it has no poles in $B_r(0)$  {(since we ensured $R_n>r$)}. Also, $\| P_n^S(z) \| $ converges to  a constant as $n\to\infty$ because $a_{w_n}(z) \to 1$ as $|w_n| \to \infty$. Finally we can obtain a bound for \eqref{equ:bounded_P}  {over all $p=q,q+1,...$} since $\sum_{n=1}^\infty \frac{1}{n^2}$ converges.

Next, we are ready to show the sequence of functions $T_\ell(z)$ is Cauchy with the supremum norm on $B_r(0)$. To see this, we can compute for $z \in B_r(0)$ and $q \le \ell < k $,
\begin{align}
\| T_\ell(z) - T_k(z) \|
\le \left\| \prod_{n = q}^\ell P_n(z) \right\|
   \left\|\prod_{n = \ell+1}^k ( P_n(z) - P_n^S(z)) \right\|
   \left\| \prod_{k+1}^\infty  P_n^S(z) \right\| .
\end{align}
%\red{(There seems to be a flaw here.  The product is supposed to start at $n = 1$.  But if you start at $n = 1$ you can't prove the Cauchy property, because the $P_n$ with small $n$ have poles in $B_r(0)$.  See earlier note on bounded sets needing to exclude the poles.)}

We have shown the first term is bounded  {over all $\ell=q,q+1,...$}, and the last term is bounded as discussed above. The middle term can be made arbitrarily small for sufficiently large $\ell$ and $k$ using the bound in \eqref{equ:quad_bound}. We conclude from this that $T_\ell(z)$ is uniformly Cauchy on $B_r(0)$, and therefore converges uniformly. This completes the proof of Claim~\ref{claim:cauchy_conv}.

{
Finally, we can show $\prod_{n=q}^\infty P_n(z)= \left(\prod_{n=q}^\ell P_n(z)\right) T_\ell(z)$ converges uniformly on $B_r(0)$ using the uniform convergence of $T_\ell(z)$ and the upper bound \eqref{equ:bounded_P}. Explicitly, supposing $C(z)$ is the function satisfying $T_\ell(z) \to C(z)$ as $\ell \to \infty$ uniformly on $B_r(0)$, we find 
\begin{align}
\left\| \prod_{n=q}^\infty  P_n(z)- \left( \prod_{n=q}^\ell P_n(z) \right)C(z) \right\|
\le 
\left\| \prod_{n=q}^\ell  P_n(z) \right\| \|C(z) - T_\ell(z) \|.
\end{align}
Since $q$ was just a finite integer, we conclude that $\prod_n P_n(z)$ converges uniformly on $B_r(0)$.
}

Finally, to complete the proof of Theorem~\ref{theorem:convergence_general}, we wish to show $B(z)$ is a constant. 
To do this, it suffices to show it is a bounded function, since it is entire (using Liouville's theorem).
Since $\tilde T (z)$ is not periodic like $\tilde S(z)$, we cannot use the same proof as for  {Theorem~\ref{theorem:convergence_static}}.
We have $B(z) = \bigl( \prod_n P_n(z) \bigr)^{-1} \tilde T(z)$, and let
$B_S(z) = \bigl( \prod_n P^S_n(z) \bigr)^{-1} \tilde S(z)$.
Since $B_S = B_S(z)$ is constant  {as found in Theorem~\ref{theorem:convergence_static}}, it suffices to show $\|B_S - B_T(z)\|$ is bounded regardless of $z$.

  {We apply} part (iii) of Lemma~\ref{lemma:static_approximation}, in order to approximate $\tilde T(z)$ with $\tilde S(z)$ away from their poles $z_n$ and $w_n$.
For a small but fixed $\epsilon$ and sufficiently large $M$, if $|z| > M$ and $z \notin B_{\epsilon} (w_n)$, $z \notin B_{\epsilon} (z_n)$ 
\begin{align}
\label{equ:approx_tilde_T_with_tilde_S}
\| \tilde T(z) - \tilde S(z) \| = O(1/M).
\end{align}

 {We will next find an estimate for}  {$\|\bigl(\prod_n P_n(z)\bigr)^{-1} -\bigl( \prod^S_n P_n(z)\bigr)^{-1} \|$} for sufficiently large $|z|$ and $z \notin B_{\epsilon} (w_n),z \notin B_{\epsilon} (z_n)$, using the bounds found above.
We will show the proof for $\|\prod_n P_n(z) - \prod^S_n P_n(z)\|$ to simplify the notation,
since each $P_n(z)$ has the same form as $P_n(z)^{-1}$ (same for $P_n^S(z))$.
 {To do this, we will start with the estimate} \eqref{equ:estimate_a_terms}. The terms $(z-w_n)^{-1}$ and $(z + w_n^*)^{-1}$ can be uniformly bounded  {if we ensure that $z$ is at least $\epsilon$ away from each $w_n$, $-w_n^*$.}  {However, now we cannot bound these terms by $O(1/R_n)$ since $z$ is not confined to $B_r(0)$}.
Instead we can propagate error terms of the form $O(1/|z-w_n|)$  {and $O(1/|z + w_n^*|)$} through the analysis, obtaining an estimate similar to \eqref{equ:quad_bound},
 {
\begin{align}
\| P_n(z) - P_n^S (z) \| &= O(1/R_n^2) + O(1/(R_n |z - w_n|)) + O(1/(R_n |z + w_n^*|)).
\end{align}
}
which holds when $|w_n|,|z_n| > R_n$ for sufficiently large $R_n$.

We can continue similarly to \eqref{equ:bounded_P}, finding a sufficiently large index $q$ so that
 {
\begin{align}
\label{equ:P_P_s_bound}
\left\| \prod_{n=q}^\infty P_n(z) \right\|
\le
\left\| \prod_{n=q}^\infty P^S_n(z) \right\|
\prod_{n=q}^\infty \left( 1 + c \left(\frac{1}{n^2} + \frac{1}{n^2 | z/w_n - 1|}+ \frac{1}{n^2 |  z/w_n^* + 1|} \right)\frac{1}{\|P_n^S(z)\|} \right),
\end{align}
}
for a constant $c$.
 {
Next we examine individual terms above in \eqref{equ:P_P_s_bound} and check they are all bounded uniformly when $z \notin B_{\epsilon} (w_n), z \notin B_{\epsilon} (-w_n^*)$.
In this domain} we can obtain a bound for $\left\| \prod_{n=q}^\infty P^S_n(z) \right\|$ (which depends on our choice of $\epsilon$).  This can be done by considering $\prod_{n=1}^\infty P^S_n(z),$ which is periodic in the imaginary direction.  {The product also converges as $\Re(z) \to \pm \infty$, so we conclude $\left\| \prod_{n=q}^\infty P^S_n(z) \right\|$ is bounded uniformly for $z \notin B_{\epsilon} (w_n),z \notin B_{\epsilon} (z_n)$.} 
 { We also have $\|P_n^S(z)\|$ are uniformly bounded away from zero when $z \notin B_{\epsilon} (w_n),z \notin B_{\epsilon} (-w_n^*)$. As before,} $\sum_{n=1}^\infty \frac{1}{n^2}$ converges.  {The terms  $\sum_{n =1}^\infty \frac{1}{n^2 |z/w_n-1|}$ and  $\sum_{n=1}^\infty \frac{1}{n^2 |z/w_n^*+1|}$ are bounded uniformly when $z$ is bounded away by the fixed $\epsilon$ away from $w_n$ (and $-w_n^*$). We can combine these observations to show that} $\left\| \prod_n P_n(z) - \prod_n P_n^S(z)\right\|$ is bounded 
as long as $z \notin B_{\epsilon} (w_n),z \notin B_{\epsilon} (z_n)$ (further, the same holds for $\| \left(\prod_n P_n(z)\right) ^{-1} - \left(\prod_n P_n^S(z)\right)^{-1}\|$). 
 With the same conditions on $z$, since $\prod_n P^S_n (z)$ is also bounded, we find that $\prod_n P_n (z)$ is bounded as well.

Combining the above result with Eq.~(\ref{equ:approx_tilde_T_with_tilde_S}), 
and the boundedness of $\tilde S(z)$, we find that for sufficiently large $|z|$ for which  $z \notin B_{\epsilon} (w_n),z \notin B_{\epsilon} (-w_n^*),z \notin B_{\epsilon} (z_n)$,
\begin{align}
&\|B(z) - B_S(z) \| \nonumber \\
& \le \biggl\| \Bigl(\prod_n P_n(z) \Bigr)^{-1}\biggr\| \left\| \tilde{T}(z) -\tilde S(z)\right\|
+ \left\| \tilde{S}(z) \right\| \biggl\| \Bigl(\prod_n P_n(z) \Bigr)^{-1} - \Bigl(\prod_n P^S_n(z) \Bigr)^{-1} \biggr \|.
\end{align}
Since we can make $\epsilon$ as small as we wish, we can ensure it is small enough so that there is a sequence of discs $B_{r_n}(0)$ with $r_n \to \infty $ as $n\to\infty$ such that none of the boundaries $\partial B_{r_n} (0)$ contain points in any of the $\epsilon$ neighborhoods of the balls $B_{\epsilon} (w_n),B_{\epsilon} (-w_n^*), B_{\epsilon} (z_n)$. By the maximum modulus principle, the maximum value of $\| B(z) \|$ on each $B_{r_n}(0)$ is attained on the boundary. Further, this value is bounded independently of $z$. This implies $B(z)$ is everywhere bounded, and hence a constant (call it $B$). 
$\blacksquare$

\bibliographystyle{plain}
\bibliography{Bibliography.bib}

\end{document}